\documentclass[aps,prl,twocolumn,superscriptaddress,longbibliography,nofootinbib,secnumarabic]{revtex4-2}
\usepackage[T1]{fontenc}
\usepackage[utf8]{inputenc}
\usepackage[english]{babel}   
\usepackage{lmodern}
\usepackage{amsmath,amssymb,amsthm,bm,bbm}
\usepackage{siunitx}
\usepackage{xspace}
\usepackage{graphicx}
\usepackage{hyperref}
\usepackage{braket}
\usepackage{physics}        
\usepackage{enumitem}       
\usepackage{times}
\usepackage{siunitx}  
\usepackage{listings}
\usepackage{xcolor} 
\usepackage{mathtools }
\usepackage{tabularx}
\usepackage{tabularray}
\usepackage{threeparttable}
\usepackage{orcidlink} 

\newcommand{\rem}[1]{{\color{black}#1}}
\hypersetup{colorlinks=true,linkcolor=blue,citecolor=blue,urlcolor=blue}

\newcommand{\Diss}[1]{\mathcal{D}[#1]}
\newcommand{\ii}{\mathrm{i}}

\begin{document}

\title{Autonomous phonon maser in levitated spin--mechanics}

\author{Mohamed Hatifi\,\orcidlink{0009-0005-3368-2751}}

\affiliation{Aix Marseille Univ, CNRS, Centrale M\'editerran\'ee, Institut Fresnel, Marseille, France}
\email{hatifi@fresnel.fr}

\begin{abstract}
Levitated nanodiamonds hosting a single nitrogen--vacancy (NV) center provide an ultra-low-frequency mechanical mode with widely tunable dissipation and spin backaction under microwave dressing and optical pumping. We demonstrate that the driven NV spin can be tuned to act as an inverted gain medium for the center-of-mass motion, thereby stabilizing an autonomous phonon maser. In the separation-of-timescales regime where spin dynamics is fast, adiabatic elimination yields a reduced mechanical master equation with closed-form, detuning-dependent transition rates and a sharp threshold given by the sign change of the phonon-number damping. For representative levitated-NV parameters, we find that a percent-level dressed-basis inversion is sufficient to reach the threshold, and the small-signal gain can exceed the intrinsic mechanical loss by orders of magnitude. Full master-equation simulations confirm above-threshold self-oscillation and a phase-diffusing, coherent steady state, whose saturation follows the Maxwell-Bloch prediction.
\end{abstract}

\maketitle
\paragraph{Introduction.—}
Self-sustained mechanical oscillations generated by internal gain—phonon lasing/masing—arise when a mechanical mode crosses from net damping to net amplification and develops an autonomous oscillation with phase coherence limited by diffusion \cite{kippenberg2008,vahala2010,grudinin2010,navarrourrios2015,navarrourrios2016,pettit2019,guha2020,eremeev2020,mercade2021}. Such regimes enable low-noise mechanical oscillators \cite{eremeev2020,li2021,hoj2024}, enhance force sensing and metrology \cite{guha2020,riviere2022,pellet2023,pan2024, kani2025a}, and provide a controlled setting for nonlinear dynamics in mesoscopic systems \cite{im2022,kuang2023,xiao2024}. In solid-state platforms, gain is commonly engineered by optomechanical dynamical backaction or by coupling mechanics to a driven multilevel system in a cavity \cite{vahala2010,gullans2015,navarrourrios2016,mercade2021,behrle2023,pan2024, kani2025}, most often in the MHz--GHz range where thermal noise is comparatively weak and optical or microwave fields provide direct control and readout \cite{guha2020,eremeev2020,li2021,xiao2024}. A distinct route is to seek phonon-maser action in mechanical systems valued for extreme isolation and very high quality factor—most notably levitated objects—while using a \emph{minimal microscopic} gain medium \cite{eremeev2020,tian2024,hoj2024}. Levitated nanodiamonds can host nearly free center-of-mass (COM) motion, but typical trap frequencies lie in the $\sim 10$--$10^{2}\,$Hz range, where the room-temperature thermal occupation is enormous \cite{cirio2012a,hsu2016,riviere2022,pellet2023,jin2024}. In this regime, establishing a phonon-maser \emph{phase} is not captured by generic ``negative damping'' arguments alone: one must derive a controlled reduced description from a microscopic driven--dissipative model, identify a threshold in terms of experimentally accessible rates, and show gain saturation to a stable above-threshold state with a genuine coherent component \cite{marquardt2006,kippenberg2008,ludwig2008,clerk2010,poot2012,lorch2014,aspelmeyer2014}. Equally important, one must formulate an operational visibility criterion in the presence of a large incoherent background \cite{hossein-zadeh2006,grudinin2010,fong2014,xiong2023}; while self-oscillation can exist even for $n_{\rm th}\!\gg\!1$, unambiguous detection typically relies on a resolvable narrow spectral feature or linewidth collapse, or on a reduced effective COM temperature achieved by feedback cooling \cite{cohadon1999,gieseler2012,guo2019}.

Levitated nanodiamonds hosting a single nitrogen--vacancy (NV) center provide a compelling setting for this program \cite{neukirch2015,pettit2017,conangla2018}. The NV spin can be coherently driven by microwaves and optically repumped \cite{doherty2013}, and magnetic-gradient coupling converts COM displacement into a controllable spin--phonon interaction \cite{kolkowitz2012,yin2013,perdriat2021a}. Phonon lasing with NV centers has been explored for clamped resonators and related cavity-based settings \cite{kepesidis2013,giannelli2016}, but the levitated COM regime combines ultra-low frequency with exceptionally weak intrinsic dissipation \cite{millen2020,gonzalez-ballestero2021,dania2024} and admits a unified microscopic description in which the same control levers interpolate between dissipative thermodynamic operation and coherent energy exchange. In particular, microwave dressing and optical repumping in the levitated-NV platform have been used to implement mechanical cooling, heat-engine operation, and coherent work storage \cite{hatifi2025}. We investigate here whether a single microwave-driven and optically repumped NV center can produce net negative phonon-number damping and stabilize COM phonon masing in a levitated oscillator despite the extreme GHz--Hz frequency mismatch \cite{hartmann1962}, and we identify the corresponding visibility condition.

We show that, in the microwave-dressed basis near resonance with the COM mode, the driven NV acts as a tunable nonequilibrium reservoir whose backaction can enhance damping or produce gain, with the sign determined by the dressed-state inversion. In the separation-of-timescales regime where spin relaxation is fast compared with mechanical dynamics, adiabatic elimination yields a reduced mechanical master equation with closed-form, detuning-dependent transition rates \cite{gardiner2010}. This provides an analytic maser threshold from the sign change of the phonon-number damping and a Maxwell--Bloch saturation law for the above-threshold limit-cycle intensity, corroborated by numerical simulations, together with a conservative window for resolving the maser signature against the thermal background \cite{scully1967,clerk2010}. More broadly, the results connect microscopic spin control to a macroscopic nonequilibrium instability in an ultra-low-frequency, high-$Q$ levitated platform, opening a route to tunable autonomous oscillators and driven--dissipative physics in regimes where thermal fluctuations would otherwise dominate.
\paragraph{Model.—} We consider a levitated nanodiamond of mass $m$ whose center-of-mass motion along the trapping axis $\Vec{e}_z$ is accurately captured by a single harmonic phonon mode of frequency $\omega_m$ with annihilation (resp. creation) operator $a$ (resp. $a^\dagger$) \cite{bullier2020,rusconi2022, perdriat2024, hatifi2025},
\begin{equation}
H_m=\hbar\omega_m a^\dagger a .
\label{eq:Hm}
\end{equation}
\begin{figure}[t]
    \centering
    \includegraphics[width=0.99\linewidth]{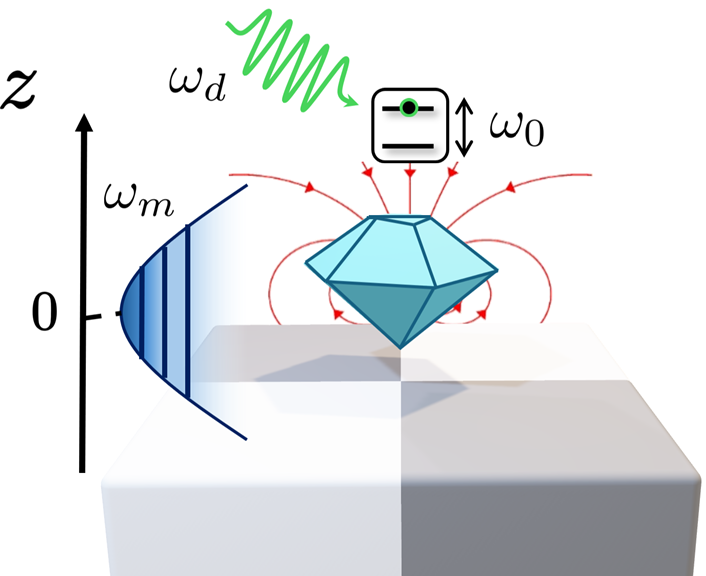}
   \caption{\textbf{Schematic of the levitated NV spin--mechanics platform.}
A nanodiamond hosting a single NV center is levitated above a source of static magnetic-field gradient (red field lines), coupling the spin to the center-of-mass coordinate $z$. The COM motion along the trap axis is approximated as a harmonic oscillator of frequency $\omega_m$. A microwave field at frequency $\omega_d$ drives a selected effective two-level transition of bare splitting $\omega_0$ and dresses the spin; when the dressed splitting $\tilde\omega \simeq \omega_m$, the interaction becomes near resonant and enables coherent exchange between spin excitations and phonons.}

    \label{fig:LeviNV}
\end{figure}
A single negatively charged NV center in the nanodiamond provides the microscopic gain medium. Its ground state is a spin triplet $\{|m_s=0\rangle,|m_s=\pm1\rangle\}$. A static magnetic field aligned with the NV axis $B_0\,\Vec{e}_z$ lifts the $\pm1$ degeneracy, allowing a near-resonant microwave drive to address one transition selectively. We restrict to the two-level manifold $\ket{0}\equiv \ket{m_s=0}$ and $\ket{1}\equiv \ket{m_s=-1}$ and neglect $\ket{m_s=+1}$ as far detuned. We note \(\omega_0=D-\gamma B_0\), the effective energy transition $\ket{0}\!\leftrightarrow\!\ket{1}$ where $D$ is the zero-field splitting of the electronic energy levels and  $\gamma\sim 28 \,{\rm GHz/Tesla}$ is the gyromagnetic ratio of the electron \cite{doherty2013}. In a frame rotating at the microwave frequency $\omega_d$ and within the rotating-wave approximation for the drive \cite{scully1997}, the driven two-level Hamiltonian reads
\begin{equation}
H_s=\frac{\hbar\Delta}{2}\sigma_z+\frac{\hbar\Omega}{2}\sigma_x,
\qquad
\Delta=\omega_0-\omega_d,
\label{eq:Hs}
\end{equation}
where $\Omega$ is the Rabi frequency, and $\sigma_z=\ket{1}\!\bra{1}-\ket{0}\!\bra{0}$. A magnetic-field gradient \cite{rabl2009,kolkowitz2012} along the mechanical coordinate $\Vec{e}_z$ produces (to first order) only a position-dependent Zeeman shift of the state $\ket{1}$ since the spin energy state $|m_s=0\rangle$ is independent of the magnetic field. Expanding the total magnetic field $B(z)=B_0+Gz$ and writing $z=z_{\rm zpf}(a+a^\dagger)$, where $z_{\text{zpf}}=\sqrt{\hbar/(2m\omega_m)}$ is the zero point fluctuation length, yields the spin--mechanical coupling
\begin{equation}
H_{sm}=\hbar\lambda (a+a^\dagger)\ket{1}\!\bra{1}.
\label{eq:Hsm_proj}
\end{equation}
where  $\lambda=\gamma G \,z_{\text{zpf}}$. Using $\ket{1}\!\bra{1}=(\mathbf{1}+\sigma_z)/2$, Eq.~\eqref{eq:Hsm_proj} separates into a spin-independent constant force and a longitudinal coupling. The constant force is absorbed by a static displacement of the mechanical mode, leaving 
\begin{equation}
H_{sm}=\hbar g_0\,(a+a^\dagger)\sigma_z,
\label{eq:Hsm}
\end{equation}
with $g_0\equiv \lambda/2$. To make the near-resonant exchange process explicit, we diagonalize $H_s$ in the dressed basis $\{\ket{\uparrow},\ket{\downarrow}\}$. Defining the dressed splitting $\tilde\omega=\sqrt{\Delta^2+\Omega^2}$, and the mixing angle $\cos\theta=\Delta/\tilde\omega$, $\sin\theta=\Omega/\tilde\omega$, the rotation into the dressed basis gives \cite{delord2020,hatifi2025}
\begin{equation}
H_{sm}=\hbar g_0(a+a^\dagger)\bigl(\cos\theta\,\tilde\sigma_z+\sin\theta\,\tilde\sigma_x\bigr),
\label{eq:Hsm_dressed}
\end{equation}
with $\tilde\sigma_z=|\uparrow\rangle\langle\uparrow|-|\downarrow\rangle\langle\downarrow|$ and $\tilde \sigma_x=|\uparrow\rangle\langle\downarrow|+|\downarrow\rangle\langle\uparrow|$.  When the dressed splitting is tuned close to the mechanical frequency, $\delta\equiv \tilde\omega-\omega_m$, and in the near-resonant weak-coupling regime ($g\ll \omega_m,\tilde\omega$ and $|\delta|\ll \omega_m$), the slowly rotating component of the transverse term yields the effective Jaynes--Cummings Hamiltonian \cite{hatifi2022a}(for more details, see Appendix)
\begin{equation}
H_{\rm JC}=\hbar\omega_m a^\dagger a+\frac{\hbar\tilde\omega}{2}\tilde\sigma_z
+\hbar g\left(a\,\tilde\sigma_+ + a^\dagger\tilde\sigma_-\right),
\label{eq:HJC}
\end{equation}
where $g\equiv g_0\sin\theta=g_0\,\Omega/\tilde\omega$. \rem{Physically, the center-of-mass motion in the harmonic trap modulates the local Zeeman splitting through the magnetic-field gradient, so that displacement about the trap equilibrium acts as a state-dependent force in the bare basis; after microwave dressing, only the transverse component of this coupling mediates near-resonant spin--phonon exchange.} Dissipation is described by a Markovian master equation for the joint density operator $\rho$,
\begin{equation}
\dot\rho=-\frac{\ii}{\hbar}[H_{\rm JC},\rho]+\mathcal{L}_m\rho+\mathcal{L}_s\rho .
\label{eq:ME}
\end{equation}
The mechanical Liouvillian is
\begin{equation}
\mathcal{L}_m\rho=\gamma_m(\bar n_{\rm th}+1)\Diss{a}\rho+\gamma_m\bar n_{\rm th}\Diss{a^\dagger}\rho,
\label{eq:Lm}
\end{equation}
with $\Diss{O}\rho\equiv O\rho O^\dagger-\tfrac12\{O^\dagger O,\rho\}$. We note  $\bar{n}_{\rm{th}}=(\exp[\hbar\omega_m/k_BT]-1)^{-1}$ the mean thermal occupation of the mechanical mode at temperature $T$ of the thermal environments, and $k_B$ the Boltzmann constant \cite{breuer2002}. The driven and optically pumped spin is modeled \emph{in the dressed basis} by the effective Liouvillian
\begin{equation}
\mathcal{L}_s\rho=
\gamma_\downarrow\Diss{\tilde\sigma_-}\rho+\gamma_\uparrow\Diss{\tilde\sigma_+}\rho
+\frac{\gamma_\phi}{2}\Diss{\tilde\sigma_z}\rho,
\label{eq:Ls}
\end{equation}
\begin{figure*}[t]
    \centering
    \includegraphics[width=.95\textwidth]{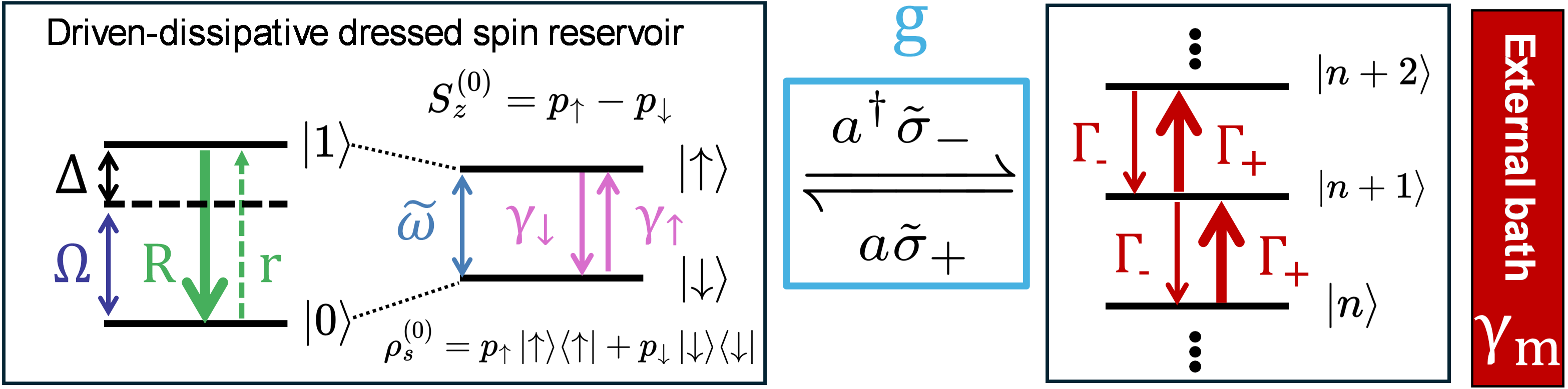}
     \caption{\rem{\textbf{Schematic mechanism of phonon gain.} A coherently driven bare spin, with microwave parameters $\Omega$ and $\Delta$ and optical recycling rates $R$ and $r$, is mapped to a driven-dissipative dressed-spin reservoir with stationary populations $p_\uparrow$ and $p_\downarrow$ and inversion $S_z^{(0)}=p_\uparrow-p_\downarrow$. Near resonance, the dressed spin exchanges excitations with the mechanical mode through $a^\dagger\tilde{\sigma}_-$ and $a\tilde{\sigma}_+$. After adiabatic elimination of the fast spin dynamics, this yields effective upward and downward phonon transitions at rates $\Gamma_+$ and $\Gamma_-$, competing with mechanical damping $\gamma_m$. In the gain regime, $\Gamma_+>\Gamma_-$, so the phonon ladder is climbed until the maser amplitude is saturated.}}
    \label{fig:maser_schematic}
\end{figure*}
\begin{figure*}[t]
    \centering
    \includegraphics[width=.9\textwidth]{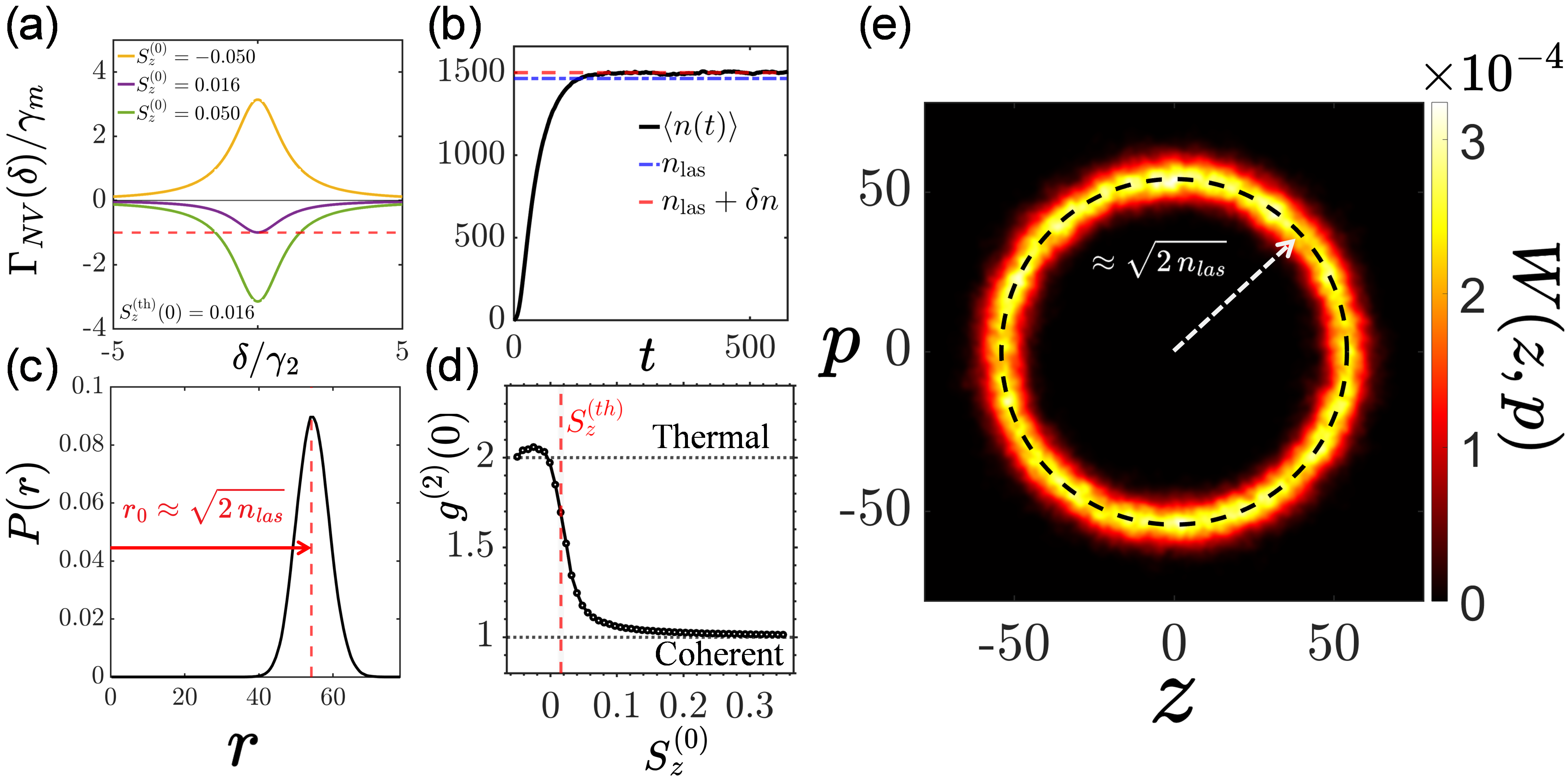}
    \caption{\rem{\textbf{Gain, saturation, and phase-diffusing steady state of the levitated-NV phonon maser.}
    \textbf{(a)} Spin-induced contribution to the phonon-number damping,
    $\Gamma_{\mathrm{NV}}(\delta)\equiv\Gamma_-(\delta)-\Gamma_+(\delta)$, normalized to the intrinsic loss $\gamma_m$, as a function of detuning $\delta/\gamma_2$ for several dressed inversions $S_z^{(0)}$. The Lorentzian response changes sign with $S_z^{(0)}$; the horizontal dashed line marks threshold, $\Gamma_{\mathrm{NV}}(0)=-\gamma_m$, corresponding to $S_z^{(\mathrm{th})}=0.016$ for the parameters used.
    \textbf{(b)} Langevin evolution of the mean phonon occupation $\langle n(t)\rangle$, showing exponential growth above threshold followed by saturation. The blue dashed line denotes the Maxwell--Bloch limit-cycle intensity $n_{\rm las}$ [Eq.~\eqref{app:nlas0}], while the red dashed line gives the noise-corrected stationary mean from the associated Fokker--Planck theory, $\langle n\rangle_{\rm FP}=n_{\rm las}+\delta n$, with $\delta n=(\mathrm{Var}(z)+\mathrm{Var}(p))/2$.
    \textbf{(c)} Stationary radial distribution $P(r)$ of the phase-space amplitude $r=\sqrt{z^2+p^2}=\sqrt{2n}$, peaked near the predicted limit-cycle radius $r_0\simeq\sqrt{2n_{\rm las}}$ (vertical dashed line).
    \textbf{(d)} Equal-time intensity correlations,
    $g^{(2)}(0)=(\langle n^2\rangle-\langle n\rangle)/\langle n\rangle^2$, showing the crossover from thermal bunching below threshold, $g^{(2)}(0)\approx2$, to near-coherent statistics above threshold, $g^{(2)}(0)\to1$; the vertical dashed line marks $S_z^{(\mathrm{th})}$.
    \textbf{(e)} Steady-state phase-space quasiprobability $W(z,p)$ reconstructed from the semiclassical Langevin dynamics associated with the reduced mechanical master equation~\eqref{eq:MEred}. Above threshold the maser approaches a limit cycle with a well-defined amplitude and freely diffusing phase, giving a phase-averaged ring in phase space \cite{zhang2024a}. The dashed circle indicates the Maxwell--Bloch prediction $r_0\simeq\sqrt{2n_{\rm las}}$; the finite ring width reflects amplitude fluctuations from the mechanical bath and spin-induced diffusion.
    Unless otherwise stated, parameters are those of Table~\ref{tab:est_inputs}, with $g=\eta\omega_m$, $\eta=0.1$, and effective pre-cooled occupation $\bar n_{\rm eff}=5$; panels \textbf{(b)}--\textbf{(e)} are evaluated at resonance, $\delta=\tilde\omega-\omega_m=0$, and for $S_z^{(0)}=0.2$.}}
    \label{fig2}
    \label{fig:Ring}
\end{figure*}
so that $\gamma_1=\gamma_\uparrow+\gamma_\downarrow$ and $\gamma_2=\gamma_1/2+\gamma_\phi$. \rem{In the present effective two-level description, the optical field is modeled as a reset process that preferentially transfers population from \(\ket{1}\) to \(\ket{0}\) at rate \(R\), with a weaker reverse leakage rate \(r\). The optical field intensity enters primarily through the reset rate \(R\); in the weak-pump regime one expects \(R\propto I\), while more generally \(R\) should saturate with increasing pump power. The microwave drive mixes the bare states \(\ket{0}\) and \(\ket{1}\) into the dressed states \(\ket{\uparrow}\) and \(\ket{\downarrow}\). Because of this mixing, the optical reset does not populate a single dressed eigenstate directly. Instead, it induces effective dressed-basis pumping and decay processes with rates \(\gamma_\uparrow\) and \(\gamma_\downarrow\), together with an additional dressed-basis dephasing rate \(\gamma_\phi\).

A key point is that these effective rates are explicit functions of the dressing angle \(\theta\). In particular, the microwave drive controls the dressed-state composition, while the imbalance between the bare reset and leakage rates, \(R-r\), controls the dissipative bias. The stationary dressed-state inversion
\[
S_z^{(0)}= \frac{\gamma_\uparrow-\gamma_\downarrow}{\gamma_\uparrow+\gamma_\downarrow}=
-\frac{2(R-r)\cos\theta}
{(R+r)\left(1+\cos^2\theta\right)}
\]
therefore becomes a directly tunable quantity (for more details see Appendix~\ref{app:frameME_microscopic_rates}). This shows how microwave dressing and optical pumping jointly control the magnitude and the sign of the inversion. The dephasing term \(\gamma_\phi\) accounts for additional dressed-basis decoherence beyond population relaxation, so that \(\gamma_2\) governs the decay of \(\langle \tilde{\sigma}_- \rangle\). All frame transformations are implemented directly at the level of the Lindblad generator. In the absence of spin--mechanical coupling (\(g=0\)), the stationary dressed-spin state is diagonal in the dressed basis,
\[
\rho_s^{(0)} = p_\uparrow \ket{\uparrow}\!\bra{\uparrow} + p_\downarrow \ket{\downarrow}\!\bra{\downarrow},
\]
with \(p_\uparrow=\gamma_\uparrow/\gamma_1\), \(p_\downarrow=\gamma_\downarrow/\gamma_1\), and \(\gamma_1=\gamma_\uparrow+\gamma_\downarrow\). As summarized schematically in Fig.~\ref{fig:maser_schematic}, microwave dressing converts the longitudinal gradient coupling into a near-resonant transverse exchange channel in the dressed basis, while optical pumping sets a tunable dressed-state inversion that determines the direction of the mechanical backaction derived below.}
\paragraph{Adiabatic elimination and phonon gain.—}
We work in a controlled separation-of-timescales regime, $\gamma_1,\gamma_2\gg g$ and $\gamma_1,\gamma_2\gg \gamma_m$,
so that the driven--dissipative spin relaxes to its local stationary state on a timescale much shorter than the mechanical evolution. Adiabatically eliminating the spin to second order in $g$ yields a reduced Markovian master equation for the mechanical state $\rho_m\equiv \Tr_s\rho$ \cite{gardiner2010},
\begin{align}\small
&\dot\rho_m
=
-\ii\left[(\omega_m+\delta\omega)\,a^\dagger a,\rho_m\right]\label{eq:MEred}\\
&+\left[\gamma_m(\bar n_{\rm th}+1)+\Gamma_-(\delta)\right]\Diss{a}\rho_m+\left[\gamma_m\bar n_{\rm th}+\Gamma_+(\delta)\right]\Diss{a^\dagger}\rho_m .\nonumber
\end{align}
Here $\Gamma_-(\delta)$ and $\Gamma_+(\delta)$ are the spin-induced \emph{downward} and \emph{upward} phonon transition rates, respectively, while $\delta\omega$ is a Lamb-shift renormalization. Evaluating the required stationary spin correlators in the dressed steady state yields the closed-form Lorentzian rates
\begin{equation}
\Gamma_-(\delta)=2g^2\,p_\downarrow \,\frac{\gamma_2}{\gamma_2^2+\delta^2},
\quad
\Gamma_+(\delta)=2g^2\,p_\uparrow\,\frac{\gamma_2}{\gamma_2^2+\delta^2}
\label{eq:Gammapm}
\end{equation}
and gives the expression of the Lamb-shift $\vert \delta \omega \vert=g^2\, \vert\delta \vert \,  \vert S_z^{(0)} \vert/\left(\gamma_2^2+\delta^2\right)\ll 1.$
Equation~\eqref{eq:MEred} thus identifies the dressed spin as a tunable nonequilibrium reservoir for the mechanics: it induces both upward and downward phonon transitions, and their \emph{imbalance} controls whether the net mechanical damping is enhanced (cooling) or reduced (gain). Indeed, for the phonon number $n\equiv\langle a^\dagger a\rangle$, one obtains
\begin{equation}
\dot n= -\gamma_{\rm eff}^{(n)}(\delta)\,n+\gamma_m\bar n_{\rm th}+\Gamma_+(\delta),
\label{eq:ndot}
\end{equation}
where we defined $\gamma_{\rm eff}^{(n)}(\delta)\equiv \gamma_m+\Gamma_-(\delta)-\Gamma_+(\delta)$. It is also natural to define the spin-induced contribution to the \emph{phonon-number damping},
\begin{equation}
\Gamma_{\mathrm{NV}}(\delta)\equiv \Gamma_-(\delta)-\Gamma_+(\delta)
=-\frac{2g^2\gamma_2}{\gamma_2^2+\delta^2}\,S_z^{(0)} .
\label{eq:Gammaopt}
\end{equation}
Positive dressed inversion $S_z^{(0)}>0$ implies $\Gamma_{\mathrm{NV}}(\delta)<0$, i.e.\ the spin \emph{reduces} the net damping and can render it negative (small-signal gain), whereas $S_z^{(0)}<0$ yields additional damping and cooling. In the linear regime, the same condition corresponds to the onset of exponential growth of the coherent amplitude $\langle a\rangle$ (at rate $\gamma_{\rm eff}^{(n)}/2$). The maser threshold corresponds to the change of sign of the linear number damping \cite{scully1967}, $\gamma_{\rm eff}^{(n)}(\delta)=0$, giving the threshold inversion
\begin{equation}
S_z^{(\mathrm{th})}(\delta)=\frac{\gamma_m}{2g^2}\,\frac{\gamma_2^2+\delta^2}{\gamma_2},
\label{eq:Szth}
\end{equation}
Above threshold, gain saturates due to inversion depletion. A Maxwell--Bloch treatment yields the coherent (limit-cycle) phonon population
\begin{equation}
n_{\mathrm{las}}
= n_{\mathrm{sat}}\left(\frac{S_z^{(0)}}{S_z^{(\mathrm{th})}}-1\right),
\qquad
n_{\mathrm{sat}}\equiv \frac{\gamma_1\gamma_2}{4g^2}.
\label{app:nlas0}
\end{equation}
A fully explicit derivation of Eqs.~\eqref{eq:MEred}--\eqref{app:nlas0}, including sign conventions and the interaction-picture bookkeeping, is given in the Appendix.
\paragraph{Semiclassical phase-space picture and ring steady state.—} Full master-equation simulations provide a microscopic benchmark of the maser regime, but they become numerically demanding once the mechanical steady state requires large Fock-space truncations. To obtain physically transparent phase-space diagnostics in that regime we therefore introduce a semiclassical drift--diffusion description consistent with the reduced mechanical master equation~\eqref{eq:MEred}; the detailed construction is also given in the Appendix. Rewriting \eqref{eq:MEred} in the following
\(
\dot\rho_m=-i[(\omega_m+\delta\omega)a^\dagger a,\rho_m]
+A_-\,\mathcal D[a]\rho_m+A_+\,\mathcal D[a^\dagger]\rho_m
\)
with total downward/upward rates \(A_-\) and \(A_+\), the associated Wigner function \(W(z,p,t)\) obeys a Fokker--Planck equation with linear drift and isotropic diffusion for the dimensionless quadratures \(z=(a+a^\dagger)/\sqrt2\) and \(p=(a-a^\dagger)/(i\sqrt2)\),
\begin{equation}
\begin{aligned}
\partial_t W={}&-\partial_z\!\big[(\mu \,z+\Omega_m p)W\big]
-\partial_p\!\big[(\mu\, p-\Omega_m z)W\big] \\
&\quad + D_{\rm W}(\partial_z^2+\partial_p^2)W ,
\end{aligned}
\label{eq:FP_main}
\end{equation}
where \(\Omega_m\simeq\omega_m+\delta\omega\) is the residual rotation frequency in phase space and the diffusion constant is fixed by the total jump rates,
\begin{equation}
D_{\rm W}=\frac{A_-+A_+}{4}
=\frac{\gamma_m(2\bar n_{\rm eff}+1)+\Gamma_-(\delta)+\Gamma_+(\delta)}{4}.
\label{eq:DW_main}
\end{equation}
%
A purely linear theory would predict a Gaussian steady state, but in the maser regime the gain saturates because stimulated emission depletes the dressed inversion. Incorporating the Maxwell--Bloch saturation law amounts to promoting the drift to an amplitude-dependent form,
\begin{equation}
\mu(n)= -\frac{\gamma_m}{2}+G(\delta)\,\frac{S_z^{(0)}}{1+n/n_{\rm sat}},
\quad
G(\delta)\equiv \frac{g^2\gamma_2}{\gamma_2^2+\delta^2},
\label{eq:mu_main}
\end{equation}
with \(n\simeq(z^2+p^2)/2\) in the large-occupation regime and \(n_{\rm sat}\) the saturation scale. Eq.~\eqref{eq:FP_main} is equivalent to the It\^o Langevin system \cite{gardiner2010}:
\begin{align}
dz &= (\mu(n)\,z+\Omega_m\, p)\,dt+\sqrt{2D_{\rm W}}\,d\xi_z,\nonumber\\
dp &= (\mu(n)\,p-\Omega_m\, z)\,dt+\sqrt{2D_{\rm W}}\,d\xi_p,
\label{eq:Langevin_main}
\end{align}
with independent Wiener noises \(d\xi_{z,p}\). The deterministic part exhibits a stable limit cycle above threshold, with radius \(r_0=\sqrt{z^2+p^2}=\sqrt{2n_{\rm las}}\) fixed by \(\mu(n_{\rm las})=0\), reproducing the semiclassical lasing intensity \(n_{\rm las}\). Physically, the amplitude is stabilized by gain clamping while the global phase remains neutrally stable; the isotropic diffusion in Eq.~\eqref{eq:Langevin_main}, therefore, produces phase diffusion and a uniform stationary distribution in angle but concentrated near \(r_0\), i.e.\ a ring in phase space which can be clearly observed in Fig. \ref{fig2}(c) and Fig. \ref{fig:Ring}.

Because the drift and diffusion are rotationally symmetric (up to the trivial rotation at \(\Omega_m\)), the steady-state solution can be written in closed form in terms of the radial coordinate \(r\):
\begin{equation}
P_r(r)\propto r\,
\exp\!\left[\frac{1}{D_{\rm W}}\int_0^{r}\mu\!\bigl(n(s)\bigr)\,s\,ds\right],
\quad n(s)\simeq s^2/2,
\label{eq:Pr_main}
\end{equation}
which directly explains the emergence of a peaked ring distribution above threshold. This approach also clarifies an important diagnostic point: \(n_{\rm las}\) characterizes the coherent and deterministic limit-cycle intensity, whereas the noisy steady state contains additional incoherent fluctuations, so that typically \(\langle n\rangle_{\rm ss}>n_{\rm las}\). Finally, intensity statistics follow directly from Langevin sampling via \(g^{(2)}(0)=\big(\langle n^2\rangle-\langle n\rangle\big)/\langle n\rangle^2\), capturing the standard crossover from thermal-like fluctuations below threshold to near-Poissonian statistics above threshold as the maser field becomes macroscopic \cite{scully1997,gardiner2010}.

\begin{table}[t]
\caption{Representative parameters used for the estimates (from Table~I of Ref.~\cite{hatifi2025}). Frequencies are given as angular frequencies; all dissipative rates are in ${\rm s^{-1}}$.}
\label{tab:est_inputs}
\begin{ruledtabular}
\begin{tabular}{lcc}
Quantity & Value & Unit\\
\hline
Mechanical frequency & $\omega_m=2\pi\times 50$ & ${\rm s^{-1}}$\\
Quality factor & $Q=10^{4}$ & --\\
Mechanical damping & $\gamma_m=\omega_m/Q\simeq 3.14\times 10^{-2}$ & ${\rm s^{-1}}$\\
Spin relaxation & $\gamma_1=5\times 10^{2}$ & ${\rm s^{-1}}$\\
Spin decoherence & $\gamma_2=10^{3}$ & ${\rm s^{-1}}$\\
\end{tabular}
\end{ruledtabular}
\end{table}
\begin{table}[t]
\caption{Derived gain and saturation benchmarks versus the dimensionless coupling $g/\omega_m=\eta$, using the inputs of Table~\ref{tab:est_inputs} and the on-resonance formulas.}
\label{tab:eta_summary}
\begin{ruledtabular}
\begin{tabular}{lccc}
& $\eta=0.05$ & $\eta=0.10$ & $\eta=0.25$ \\
\hline
$g$ [$\mathrm{s^{-1}}$] & $1.57\times 10^{1}$ & $3.14\times 10^{1}$ & $7.85\times 10^{1}$ \\
$S_z^{(\mathrm{th})}$ & $6.37\times 10^{-2}$ & $1.59\times 10^{-2}$ & $2.55\times 10^{-3}$ \\
$|\Gamma_{\mathrm{NV}}|_{\max}=2g^2/\gamma_2$ [$\mathrm{s^{-1}}$] &
$4.93\times 10^{-1}$ & $1.97$ & $1.23\times 10^{1}$ \\
$|\Gamma_{\mathrm{NV}}|_{\max}/\gamma_m$ & $1.57\times 10^{1}$ & $6.27\times 10^{1}$ & $3.92\times 10^{2}$ \\
$n_{\rm sat}=\gamma_1\gamma_2/(4g^2)$ & $5.07\times 10^{2}$ & $1.27\times 10^{2}$ & $2.03\times 10^{1}$ \\
\end{tabular}
\end{ruledtabular}
\end{table}

\paragraph{Discussion and experimental implications.—}
Tables~\ref{tab:est_inputs} and \ref{tab:eta_summary} place the gain and saturation laws derived above on a quantitative footing for the levitated-NV parameter window.

\noindent For the representative operating point $\omega_m=2\pi\times 50~\mathrm{s^{-1}}$ and $Q=10^{4}$ (Table~\ref{tab:est_inputs}), the intrinsic number-damping rate is $\gamma_m=\omega_m/Q\simeq 3.1\times 10^{-2}~\mathrm{s^{-1}}$, while the dressed-spin decoherence and relaxation rates are $\gamma_2\simeq 10^{3}~\mathrm{s^{-1}}$ and $\gamma_1\simeq 5\times 10^{2}~\mathrm{s^{-1}}$. In this regime, the threshold inversion $S_z^{(\mathrm{th})}(0)=\gamma_m\gamma_2/(2g^2)$ is small because the mechanical losses are extremely weak: Table~\ref{tab:eta_summary} shows that for $\eta\equiv g/\omega_m=0.10$ the maser threshold corresponds to $S_z^{(\mathrm{th})}\simeq 1.6\times 10^{-2}$ (percent-level dressed imbalance), and even $\eta=0.05$ remains in the few-percent range. The same table makes explicit that once inversion is available the maximum small-signal gain rate $|\Gamma_{\mathrm{NV}}|_{\max}=2g^2/\gamma_2$ can exceed $\gamma_m$ by one to several orders of magnitude, so that the onset of instability and its detuning dependence are controlled primarily by the dressed coherence scale $\gamma_2$ and the effective coupling $g$ rather than by mechanical dissipation. This is precisely what is visible in Fig.~\ref{fig2}(a): $\Gamma_{\mathrm{NV}}(\delta)$ is a Lorentzian in $\delta/\gamma_2$ with a sign set by $S_z^{(0)}$, and the threshold crossing corresponds to $\Gamma_{\mathrm{NV}}(0)=-\gamma_m$. Above threshold, gain clamping stabilizes the oscillation amplitude at the Maxwell--Bloch intensity $n_{\mathrm{las}}$ Eq.~\eqref{app:nlas0}, while phase diffusion prevents a unique steady-state phase. The time-domain saturation in Fig.~\ref{fig2}(b) and the peaked radial distribution in Fig.~\ref{fig2}(c) are therefore the natural dynamical and phase-space signatures of the same mechanism: deterministic attraction to a limit-cycle intensity, combined with angular diffusion. The corresponding phase-space picture is made explicit by the Wigner reconstruction in Fig.~\ref{fig:Ring}, where the steady state forms a ring concentrated near the predicted radius $r_0\simeq\sqrt{2 n_{\rm las}}$ and broadened by diffusion. Consistently, Fig.~\ref{fig2}(d) shows the standard threshold crossover in intensity fluctuations: below threshold the oscillator is dominated by thermal-like statistics ($g^{(2)}(0)\approx 2$), while above threshold the relative fluctuations are suppressed as the field becomes macroscopic ($g^{(2)}(0)\to 1$).

\rem{A key practical point in the levitated, ultra-low-frequency regime is that \emph{negative damping is not by itself an unambiguous lasing signature} \cite{hossein-zadeh2006, grudinin2010}: for \(\omega_m\) in the \(10\)–\(10^2\)\,Hz range, the thermal occupation at ambient temperature is enormous, so a coherent component is observable only if the effective motional occupation \(\bar n_{\rm eff}\) is sufficiently reduced. This requirement is made quantitative in Appendix, where we derive a conservative design inequality and show that, for the benchmark parameter set used here, the required \(T_{\rm eff}\) remains highly demanding for present low-frequency levitated platforms. In practice, this motivates a staged protocol in which the motion is first cooled to a low \(\bar n_{\rm eff}\) (e.g.\ using the protocol of Ref.~\cite{hatifi2025}, feedback or sideband cooling \cite{gieseler2012}), and only then the operating point is switched to the maser configuration, so that the threshold remains set by the intrinsic \(\gamma_m\) while the initial noise floor is reduced.}

\paragraph{Conclusion.—}
In this Letter, we established that a single microwave-driven and optically repumped NV center can serve as a minimal gain medium for the center-of-mass motion of a levitated nanodiamond and drive it through an autonomous phonon-masing threshold. In the controlled separation-of-timescales regime, where spin dynamics is fast compared to the mechanics, adiabatic elimination yields a reduced mechanical master equation with closed-form, detuning-dependent transition rates, resulting in a sharp threshold from the sign change of the phonon-number damping and an analytic Maxwell-Bloch saturation law. A key physical point is the microwave dressing, which bridges the extreme bare frequency mismatch \cite{hartmann1962} between the NV spin (GHz) and levitated motion (10--$10^2$ Hz) by generating a tunable dressed splitting that can be brought near $\omega_m$, enabling resonant spin--phonon exchange and gain. The resulting phase-diffusing limit cycle manifests as a ring-like phase-space steady state and a crossover of intensity statistics across threshold, providing operational diagnostics of maser action within the model.  An immediate extension is to incorporate realistic pumping and technical noise channels to predict measurable displacement spectra and phase noise-linewidth in concrete readout schemes. A natural next step is to optimize combined pre-cooling (or continuous feedback) and gain protocols to delineate the most accessible visibility window at ultra-low frequencies. More broadly, these results provide a transparent route from microscopic-driven spin control to macroscopic nonequilibrium self-oscillation in high-$Q$ mechanics, even in regimes dominated by extreme frequency mismatch and large thermal backgrounds.

\paragraph{Acknowledgements—}The author would like to thank Jason Twamley and Anshuman Nayak for insightful discussions. 
\bibliographystyle{apsrev4-2}
%

\clearpage
\setcounter{secnumdepth}{1}
\setcounter{secnumdepth}{2} 

\appendix

\section{Physical origin of the exchange interaction and derivation of the Jaynes--Cummings limit}
\label{app:JCderive}

The microscopic spin--mechanical coupling is induced by the magnetic-field gradient: the center-of-mass displacement modulates the local Zeeman field experienced by the NV electronic spin. In the $\{\ket{0},\ket{1}\}$ manifold, $\ket{0}$ is (to leading order) insensitive to the linear Zeeman shift, whereas $\ket{1}$ shifts linearly with the field. Consequently, the gradient converts displacement into a \emph{state-dependent} force on the mechanics, i.e.\ a longitudinal coupling in the bare basis. The key mechanism enabling resonant \emph{energy exchange} is that the microwave drive dresses the spin eigenstates, so that this longitudinal coupling acquires a transverse component in the dressed basis. Near resonance, that transverse component reduces to a beam-splitter interaction.

\paragraph{Starting Hamiltonian and removal of the spin-independent force.}
In the microwave rotating frame, the coherent Hamiltonian is
\begin{equation}
H = \hbar\omega_m a^\dagger a
+\frac{\hbar\Delta}{2}\sigma_z+\frac{\hbar\Omega}{2}\sigma_x
+\hbar\lambda (a+a^\dagger)\ket{1}\!\bra{1},
\label{app:H_start}
\end{equation}
where $\sigma_z=\ket{1}\!\bra{1}-\ket{0}\!\bra{0}$ and $\sigma_x=\sigma_+ + \sigma_-$. Using
$\ket{1}\!\bra{1}=(\mathbf{1}+\sigma_z)/2$, the coupling decomposes as
\begin{equation}
\hbar\lambda (a+a^\dagger)\ket{1}\!\bra{1}
=\frac{\hbar\lambda}{2}(a+a^\dagger)\mathbf{1}
+\frac{\hbar\lambda}{2}(a+a^\dagger)\sigma_z .
\label{app:split_force}
\end{equation}
The first term is a spin-independent constant force that merely shifts the mechanical equilibrium position and produces no backaction mechanism by itself. It is removed by a static displacement leaving the longitudinal coupling
\begin{equation}
H_{sm}=\hbar g_0(a+a^\dagger)\sigma_z,
\qquad g_0\equiv \lambda/2,
\label{app:Hsm_long}
\end{equation}
which is the starting point for the dressed-state analysis.

\paragraph{Dressing and emergence of a transverse component.}
We diagonalize the driven spin Hamiltonian by the rotation
$U=\exp(-i\theta\sigma_y/2)$ with
\begin{equation}
\tilde\omega=\sqrt{\Delta^2+\Omega^2},
\quad
\cos\theta=\Delta/\tilde\omega,
\quad
\sin\theta=\Omega/\tilde\omega,
\label{app:theta_def}
\end{equation}
so that
\begin{equation}
\frac{\hbar\Delta}{2}\sigma_z+\frac{\hbar\Omega}{2}\sigma_x=\frac{\hbar\tilde\omega}{2}\tilde\sigma_z,
\label{app:dressed_rel}
\end{equation}
with $\sigma_z=\cos\theta\,\tilde\sigma_z+\sin\theta\,\tilde\sigma_x$ and $\tilde\sigma_x=\tilde\sigma_+ + \tilde\sigma_- $. Substituting into Eq.~\eqref{app:Hsm_long} yields
\begin{equation}
H_{sm}
=\hbar g_0\cos\theta\,(a+a^\dagger)\tilde\sigma_z
+\hbar g\,(a+a^\dagger)(\tilde\sigma_+ + \tilde\sigma_-),
\label{app:Hsm_dressed_full}
\end{equation}
where $g\equiv g_0\sin\theta=g_0\frac{\Omega}{\tilde\omega}.$  The first term remains longitudinal in the dressed basis and therefore does not mediate resonant exchange; it acts dispersively. The second term is transverse and is responsible for near-resonant spin--phonon exchange.

\paragraph{Interaction-picture decomposition and rotating-wave approximation.}
We now move to the interaction picture generated by
\begin{equation}
H_0=\hbar\omega_m a^\dagger a+\frac{\hbar\tilde\omega}{2}\tilde\sigma_z,
\label{app:H0}
\end{equation}
so that $a(t)=ae^{-i\omega_m t}$,$a^\dagger(t)=a^\dagger e^{+i\omega_m t}$ and $\tilde\sigma_\pm(t)=\tilde\sigma_\pm e^{\pm i\tilde\omega t}$.
The transverse interaction becomes
\begin{align}
V_{\perp,I}(t)
&=\hbar g\,(a e^{-i\omega_m t}+a^\dagger e^{+i\omega_m t})
(\tilde\sigma_+ e^{+i\tilde\omega t}+\tilde\sigma_- e^{-i\tilde\omega t})
\nonumber\\
&=\hbar g\Big[
a\,\tilde\sigma_+ e^{+i(\tilde\omega-\omega_m)t}
+a^\dagger\tilde\sigma_- e^{-i(\tilde\omega-\omega_m)t}
\nonumber\\
&\qquad\quad
+a\,\tilde\sigma_- e^{-i(\tilde\omega+\omega_m)t}
+a^\dagger\tilde\sigma_+ e^{+i(\tilde\omega+\omega_m)t}
\Big].
\label{app:VperpI}
\end{align}
Defining the detuning $\delta\equiv\tilde\omega-\omega_m$, the first two terms rotate at $\pm\delta$ and are resonant when $|\delta|\ll \omega_m$. The last two terms rotate at $\pm(\tilde\omega+\omega_m)$ and correspond to simultaneous creation/annihilation of a spin excitation and a phonon; they are strongly off-resonant. In the near-resonant weak-coupling regime
\begin{equation}
|\delta|\ll \tilde\omega+\omega_m,
\qquad
g\ll \omega_m,\tilde\omega,
\label{app:RWAconds}
\end{equation}
these counter-rotating contributions average out on the timescale of the resonant exchange dynamics, leaving
\begin{equation}
V_{{\rm JC},I}(t)\simeq \hbar g\Big(a\,\tilde\sigma_+ e^{+i\delta t}+a^\dagger\tilde\sigma_- e^{-i\delta t}\Big).
\label{app:VJC_I}
\end{equation}
Returning to the Schr\"odinger picture yields the effective Jaynes--Cummings Hamiltonian
\begin{equation}
H_{\rm JC}=\hbar\omega_m a^\dagger a+\frac{\hbar\tilde\omega}{2}\tilde\sigma_z
+\hbar g\left(a\,\tilde\sigma_+ + a^\dagger\tilde\sigma_-\right),
\label{app:HJC_final}
\end{equation}
used in the main text.

\paragraph{Remarks on omitted terms.}
The dressed longitudinal coupling $g_0\cos\theta\,(a+a^\dagger)\tilde\sigma_z$ does not exchange energy between spin and mechanics; in the interaction picture it oscillates at $\omega_m$ and contributes primarily dispersive effects (equilibrium shifts and small renormalizations). Likewise, the counter-rotating terms neglected in Eq.~\eqref{app:VperpI} generate only small corrections of order $g^2/(\tilde\omega+\omega_m)$ in the regime \eqref{app:RWAconds}. The gain physics analyzed in this work is governed by the near-resonant exchange term \eqref{app:VJC_I}, whose competition with spindissipation and inversion produces the threshold behavior derived in the following.

\section{Derivation of the effective mechanical master equation and the maser steady state}
\label{app:elim}

This Appendix provides a fully explicit derivation of (i) the effective reduced dynamics of the mechanical mode obtained by eliminating the driven--dissipative dressed spin, and (ii) the semiclassical steady state above threshold (gain saturation). Throughout, we keep track of all prefactors and signs, and we state clearly the approximations under which each step is controlled. For notational simplicity, we drop the tildes on dressed spin operators within the remainder of the Appendix; all spin operators are in the dressed basis.

\subsection{Starting point, notation, and assumptions}
\label{app:setup}
We consider the driven--dissipative Jaynes--Cummings model used in the main text,
\begin{align}
H_{\mathrm{JC}}
&= \hbar \omega_m a^\dagger a + \frac{\hbar \tilde\omega}{2}\,\sigma_z
+ \hbar g \bigl(a \sigma_+ + a^\dagger \sigma_-\bigr),
\label{app:HJC}
\end{align}
where $a$ is the annihilation operator of the mechanical mode of frequency $\omega_m$, and $\sigma_{z,\pm}$ act on the \emph{dressed} spin eigenbasis $\{\ket{\uparrow},\ket{\downarrow}\}$ with $\sigma_z=\ket{\uparrow}\!\bra{\uparrow}-\ket{\downarrow}\!\bra{\downarrow}$, $\sigma_-=\ket{\downarrow}\!\bra{\uparrow}$, and $\sigma_+=\ket{\uparrow}\!\bra{\downarrow}$. The full density operator $\rho$ obeys the Markovian master equation
\begin{equation}
\dot\rho = -\frac{i}{\hbar}[H_{\mathrm{JC}},\rho] + \mathcal{L}_m\rho + \mathcal{L}_s\rho .
\label{app:MEfull}
\end{equation}
The mechanical Liouvillian is
\begin{equation}
\mathcal{L}_m\rho
= \gamma_m(\bar n_{\mathrm{th}}+1)\mathcal{D}[a]\rho
+ \gamma_m\bar n_{\mathrm{th}}\mathcal{D}[a^\dagger]\rho,
\label{app:Lm}
\end{equation}
with $\mathcal{D}[O]\rho = O\rho O^\dagger - \frac12\{O^\dagger O,\rho\}$ and $\bar{n}_{\rm{th}}=(\exp[\hbar\omega_m/k_BT]-1)^{-1}$ the mean thermal occupation of the mechanical mode at temperature $T$. The dressed spin Liouvillian is taken in the standard form
\begin{equation}
\mathcal{L}_s\rho
= \gamma_\downarrow \mathcal{D}[\sigma_-]\rho
+ \gamma_\uparrow \mathcal{D}[\sigma_+]\rho
+ \frac{\gamma_\phi}{2}\,\mathcal{D}[\sigma_z]\rho .
\label{app:Ls}
\end{equation}
With this convention, the longitudinal and transverse decay rates are
\begin{equation}
\gamma_1=\gamma_\uparrow+\gamma_\downarrow,\qquad
\gamma_2=\frac{\gamma_1}{2}+\gamma_\phi ,
\label{app:gamma12}
\end{equation}
so that, in the absence of coupling to the mechanics, $\langle \sigma_z\rangle$ relaxes at rate $\gamma_1$ and $\langle \sigma_-\rangle$ decays at rate $\gamma_2$. The stationary spin state $\rho_s^{(0)}$ (for $g=0$) is diagonal in the dressed basis,
\begin{equation}
\rho_s^{(0)} = p_\uparrow \ket{\uparrow}\!\bra{\uparrow} + p_\downarrow  \ket{\downarrow}\!\bra{\downarrow},\qquad
p_\uparrow=\frac{\gamma_\uparrow}{\gamma_1},\quad p_\downarrow =\frac{\gamma_\downarrow}{\gamma_1},
\label{app:rhos0}
\end{equation}
and its inversion is
\begin{equation}
S_z^{(0)} \equiv \Tr(\sigma_z \rho_s^{(0)}) = p_\uparrow - p_\downarrow  = \frac{\gamma_\uparrow-\gamma_\downarrow}{\gamma_1}.
\label{app:Sz0}
\end{equation}

\emph{Controlled regime.}
The adiabatic elimination below is controlled in the separation-of-timescales regime
\begin{equation}
\gamma_1,\gamma_2 \gg g,\qquad \gamma_1,\gamma_2 \gg \gamma_m,
\label{app:timescale}
\end{equation}
and for weak spin--mechanics coupling in the sense that the spin remains close to $\rho_s^{(0)}$ (up to corrections of order $g$). This is the same physical regime underlying standard sideband cooling and maser gain derivations.

\subsection{Master equation under rotating frames, spin rotations, and mechanical displacements}
\label{app:frameME}

In the main text, we use three standard changes of representation: (i) a rotating frame for the microwave drive, (ii) a rotation into the microwave-dressed spin basis, and (iii) a static displacement of the mechanical mode to absorb a spin-independent force term. Since our analysis relies on a Markovian (GKSL) master equation, it is essential to explicitly state how these transformations affect both the Hamiltonian and the dissipators.

\subsubsection{General unitary change of frame}
\label{app:frameME_general}

Consider a Markovian master equation in GKSL form,
\begin{equation}
\dot\rho=\mathcal{L}\rho
\equiv -\frac{i}{\hbar}[H,\rho]+\sum_\mu \gamma_\mu\,\mathcal{D}[L_\mu]\rho,
\quad
\mathcal{D}[L]\rho=L\rho L^\dagger-\frac12\{L^\dagger L,\rho\}.
\label{app:GKSL_general}
\end{equation}
Let $U(t)$ be a (possibly time-dependent) unitary and define the transformed state
$\tilde\rho\equiv U(t)\rho U^\dagger(t)$. A direct differentiation gives
\begin{equation}
\dot{\tilde\rho}=
-\frac{i}{\hbar}[\tilde H(t),\tilde\rho]
+\sum_\mu \gamma_\mu\,\mathcal{D}[\tilde L_\mu(t)]\tilde\rho,
\label{app:GKSL_transformed}
\end{equation}
with
\begin{equation}
\tilde L_\mu(t)=U(t)L_\mu U^\dagger(t),
\quad
\tilde H(t)=U(t)HU^\dagger(t)+i\hbar\,\dot U(t)U^\dagger(t).
\label{app:H_L_transform}
\end{equation}
Thus, \emph{the GKSL structure is preserved} under a unitary change of frame, with transformed jump operators and with an additional Hamiltonian term $i\hbar\dot U U^\dagger$ when the transformation is time dependent (as in a rotating frame). In practice, after entering a rotating frame one typically applies a rotating-wave/secular approximation to remove residual fast oscillations in $\tilde H(t)$ and in $\tilde L_\mu(t)$, yielding a time-independent generator again (this is the standard derivation of time-independent driven dissipative models).

\subsubsection{Static mechanical displacement and invariance of the thermal Liouvillian}
\label{app:frameME_displacement}
In the main text we use a static displacement of the mechanical operator to remove a spin-independent force term. Let $D(\alpha)=\exp(\alpha a^\dagger-\alpha^\ast a)$ and define $\tilde\rho=D(\alpha)\rho D^\dagger(\alpha)$, so that
\begin{equation}
D(\alpha)\,a\,D^\dagger(\alpha)=a-\alpha,
\qquad
D(\alpha)\,a^\dagger\,D^\dagger(\alpha)=a^\dagger-\alpha^\ast.
\label{app:disp_action}
\end{equation}
(Equivalently, $D^\dagger a D=a+\alpha$; the choice is a convention and only changes the sign of $\alpha$ below.) The key point is that a static displacement does not generate new dissipative physics; it only produces a coherent-drive (Hamiltonian) term. Concretely, for a constant $\alpha$ one finds the identity
\begin{equation}
\mathcal{D}[a+\alpha]\rho
=
\mathcal{D}[a]\rho
-\frac{i}{\hbar}[H_\alpha,\rho],
\qquad
H_\alpha\equiv \frac{i\hbar}{2}\left(\alpha^\ast a-\alpha a^\dagger\right).
\label{app:Dshift_identity}
\end{equation}
This follows by expanding $\mathcal{D}[a+\alpha]\rho$ and collecting the cross terms proportional to $\alpha$ and $\alpha^\ast$ into commutators; all terms proportional to $|\alpha|^2$ cancel identically because $|\alpha|^2$ is a scalar. Applying Eq.~\eqref{app:Dshift_identity} to the thermal mechanical Liouvillian \eqref{app:Lm} yields
\begin{align}
&\gamma_m(\bar n_{\rm th}+1)\mathcal{D}[a+\alpha]\rho
+\gamma_m\bar n_{\rm th}\mathcal{D}[a^\dagger+\alpha^\ast]\rho
\nonumber\\
&=
\gamma_m(\bar n_{\rm th}+1)\mathcal{D}[a]\rho
+\gamma_m\bar n_{\rm th}\mathcal{D}[a^\dagger]\rho
-\frac{i}{\hbar}[H_{\rm drv},\rho],
\label{app:Lm_displaced}
\end{align}
where the only extra term is a coherent drive
\begin{equation}
H_{\rm drv}=\gamma_m H_\alpha
=\frac{i\hbar\gamma_m}{2}\left(\alpha^\ast a-\alpha a^\dagger\right),
\label{app:Hdrv}
\end{equation}
independent of $\bar n_{\rm th}$. In other words, after a static displacement, the dissipator retains \emph{exactly} the same Lindblad form for the fluctuation operator, and one can choose $\alpha$ so that
$H_{\rm drv}$ cancels the static force term in the transformed Hamiltonian (this is the only role of the displacement used in the main text).

\subsubsection{Rotation to the dressed spin basis and the effective dressed dissipator}
\label{app:frameME_spinrotation}

The microwave drive mixes the bare two-level manifold $\{\ket{0},\ket{1}\}$, and it is convenient to work in the eigenbasis of the driven spin Hamiltonian. Let $R$ be the time-independent spin rotation that diagonalizes $H_s$, so that in the dressed basis $H_s=(\hbar\tilde\omega/2)\tilde\sigma_z$ and $\sigma_z=\cos\theta\,\tilde\sigma_z+\sin\theta\,\tilde\sigma_x$. At the level of the GKSL master equation, a time-independent rotation is implemented by $\tilde\rho=R\rho R^\dagger$ and simply maps the jump operators as
\begin{equation}
L_\mu \ \mapsto\ \tilde L_\mu = R L_\mu R^\dagger,
\label{app:jump_rotation}
\end{equation}
with no additional Hamiltonian term (since $\dot R=0$).

If one started from a microscopic dissipator written in the \emph{bare} basis, the transformed jump operators $\tilde L_\mu$ would in general contain linear combinations of $\tilde\sigma_\pm$ and $\tilde\sigma_z$. In the driven problem, however, the natural and standard procedure is to perform a secular approximation in the dressed splitting $\tilde\omega$, which removes fast-oscillating cross terms and yields a time-independent dressed-basis generator with three effective channels: downward transitions, upward transitions, and pure dephasing in the dressed basis. This is the origin of the effective dressed spin Liouvillian used in Eq.~\eqref{app:Ls},
\begin{equation}
\mathcal{L}_s\rho
=
\gamma_\downarrow \mathcal{D}[\tilde\sigma_-]\rho
+\gamma_\uparrow \mathcal{D}[\tilde\sigma_+]\rho
+\frac{\gamma_\phi}{2}\mathcal{D}[\tilde\sigma_z]\rho,
\label{app:Ls_dressed_justif}
\end{equation}
with $\gamma_{\uparrow,\downarrow,\phi}$ interpreted as \emph{effective} rates set by the combined action of optical pumping, environmental noise, and microwave dressing (not bare $T_1,T_2$ parameters in the
undriven basis). The stationary dressed state at $g=0$ is then diagonal with populations $p_\uparrow=\gamma_\uparrow/\gamma_1$ and $p_\downarrow =\gamma_\downarrow/\gamma_1$ Eq.~\eqref{app:rhos0}, which is the only property needed for the elimination carried out in the following sections.

\subsubsection{Scope of the transformation used in the main text}
\label{app:frameME_scope}

We emphasize that the displacement invoked in the main text is a \emph{static, spin-independent} displacement used solely to remove a constant force term (arising, e.g., when writing $\ket{1}\!\bra{1}=(\mathbf{1}+\sigma_z)/2$ in the truncated NV manifold). Because it is spin independent, it does not entangle spin and mechanics and does not generate additional dissipative channels beyond the coherent drive term \eqref{app:Hdrv}. The dressed-basis rotation is implemented at the level of the master equation as discussed above, and the subsequent near-resonant rotating-wave step leading to the effective Jaynes--Cummings form is treated explicitly in the interaction picture used for the adiabatic elimination.

\rem{\subsection{Microscopic origin of the effective dressed spin Liouvillian}
\label{app:frameME_microscopic_rates}

The notation becomes much cleaner if one keeps a strict distinction between the \emph{bare-basis} generator, denoted by \(\mathcal{L}_s^{(b)}\), and the \emph{effective dressed-basis} generator, denoted by \(\mathcal{L}_s\). Since, as stated at the beginning of Appendix~\ref{app:elim}, all spin operators are written in the dressed basis from that point onward, no tildes are introduced here.

We therefore start from the bare-basis dissipator
\begin{equation}
\mathcal{L}_{s}^{(b)}\rho
=
R\,\mathcal{D}[\sigma_-^{(b)}]\rho
+
r\,\mathcal{D}[\sigma_+^{(b)}]\rho
+
\frac{\gamma_{\phi,b}}{2}\,\mathcal{D}[\sigma_z^{(b)}]\rho ,
\label{app:Ls_bare_mic_consistent}
\end{equation}
where
\begin{equation}
\sigma_-^{(b)}=\ket{0}\!\bra{1},
\qquad
\sigma_+^{(b)}=\ket{1}\!\bra{0},
\qquad
\sigma_z^{(b)}=\ket{1}\!\bra{1}-\ket{0}\!\bra{0}.
\label{app:bare_ops_mic_consistent}
\end{equation}
Here \(R\) is the optical repumping rate \(\ket{1}\to\ket{0}\), \(r\) is the reverse leakage rate repopulating \(\ket{1}\), and \(\gamma_{\phi,b}\) is the bare pure-dephasing rate.

Using the same dressed basis as in Eqs.~\eqref{app:theta_def} and \eqref{app:dressed_rel}, we write
\begin{equation}
\ket{1}
=
\cos\frac{\theta}{2}\,\ket{\uparrow}
+
\sin\frac{\theta}{2}\,\ket{\downarrow},
\qquad
\ket{0}
=
-\sin\frac{\theta}{2}\,\ket{\uparrow}
+
\cos\frac{\theta}{2}\,\ket{\downarrow}.
\label{app:bare_to_dressed_states_consistent}
\end{equation}
Since all unadorned operators now refer to the dressed basis,
\begin{equation}
\sigma_-=\ket{\downarrow}\!\bra{\uparrow},
\qquad
\sigma_+=\ket{\uparrow}\!\bra{\downarrow},
\qquad
\sigma_z=\ket{\uparrow}\!\bra{\uparrow}-\ket{\downarrow}\!\bra{\downarrow}.
\label{app:dressed_ops_consistent}
\end{equation}
Substituting Eq.~\eqref{app:bare_to_dressed_states_consistent} into the bare jump operators gives
\begin{align}
\sigma_-^{(b)}
&=
\ket{0}\!\bra{1}
\nonumber\\
&=
\left(
-\sin\frac{\theta}{2}\ket{\uparrow}
+
\cos\frac{\theta}{2}\ket{\downarrow}
\right)
\left(
\cos\frac{\theta}{2}\bra{\uparrow}
+
\sin\frac{\theta}{2}\bra{\downarrow}
\right)
\nonumber\\
&=
\cos^2\frac{\theta}{2}\,\sigma_-
-
\sin^2\frac{\theta}{2}\,\sigma_+
-
\frac{\sin\theta}{2}\,\sigma_z ,
\label{app:sigmam_bare_expanded_consistent}
\\[1ex]
\sigma_+^{(b)}
&=
\ket{1}\!\bra{0}
\nonumber\\
&=
\left(
\cos\frac{\theta}{2}\ket{\uparrow}
+
\sin\frac{\theta}{2}\ket{\downarrow}
\right)
\left(
-\sin\frac{\theta}{2}\bra{\uparrow}
+
\cos\frac{\theta}{2}\bra{\downarrow}
\right)
\nonumber\\
&=
\cos^2\frac{\theta}{2}\,\sigma_+
-
\sin^2\frac{\theta}{2}\,\sigma_-
-
\frac{\sin\theta}{2}\,\sigma_z .
\label{app:sigmap_bare_expanded_consistent}
\end{align}
Likewise, Eq.~\eqref{app:dressed_rel} gives
\begin{equation}
\sigma_z^{(b)}
=
\cos\theta\,\sigma_z
+
\sin\theta\,(\sigma_+ + \sigma_-).
\label{app:sigmaz_bare_expanded_consistent}
\end{equation}
Since the dressed spin Hamiltonian is diagonal,
\begin{equation}
H_s=\frac{\hbar\tilde\omega}{2}\sigma_z,
\label{app:Hs_mic_consistent}
\end{equation}
the interaction-picture operators evolve as
\begin{equation}
\sigma_-(t)=\sigma_-e^{-i\tilde\omega t},
\qquad
\sigma_+(t)=\sigma_+e^{+i\tilde\omega t},
\qquad
\sigma_z(t)=\sigma_z .
\label{app:dressed_IP_rotation_consistent}
\end{equation}
Under the secular approximation in the dressed splitting,
\begin{equation}
\tilde\omega \gg R,\; r,\; \gamma_{\phi,b},
\label{app:secular_condition_mic_consistent}
\end{equation}
the oscillatory cross terms average out, and one obtains
\begin{align}
\mathcal{D}[\sigma_-^{(b)}]\rho
&\longrightarrow
\cos^4\frac{\theta}{2}\,\mathcal{D}[\sigma_-]\rho
+
\sin^4\frac{\theta}{2}\,\mathcal{D}[\sigma_+]\rho
+
\frac{\sin^2\theta}{4}\,\mathcal{D}[\sigma_z]\rho ,
\label{app:Dsigmam_secular_mic_consistent}
\\
\mathcal{D}[\sigma_+^{(b)}]\rho
&\longrightarrow
\cos^4\frac{\theta}{2}\,\mathcal{D}[\sigma_+]\rho
+
\sin^4\frac{\theta}{2}\,\mathcal{D}[\sigma_-]\rho
+
\frac{\sin^2\theta}{4}\,\mathcal{D}[\sigma_z]\rho ,
\label{app:Dsigmap_secular_mic_consistent}
\\
\mathcal{D}[\sigma_z^{(b)}]\rho
&\longrightarrow
\cos^2\theta\,\mathcal{D}[\sigma_z]\rho
+
\sin^2\theta\,\mathcal{D}[\sigma_+]\rho
+
\sin^2\theta\,\mathcal{D}[\sigma_-]\rho .
\label{app:Dsigmaz_secular_mic_consistent}
\end{align}
The resulting effective dressed-basis generator is therefore
\begin{equation}
\mathcal{L}_s\rho
=
\gamma_\downarrow \mathcal{D}[\sigma_-]\rho
+
\gamma_\uparrow \mathcal{D}[\sigma_+]\rho
+
\frac{\gamma_\phi}{2}\,\mathcal{D}[\sigma_z]\rho ,
\label{app:Ls_mic_consistent}
\end{equation}
with
\begin{align}
\gamma_\downarrow
&=
R\cos^4\frac{\theta}{2}
+
r\sin^4\frac{\theta}{2}
+
\frac{\gamma_{\phi,b}}{2}\sin^2\theta ,
\label{app:gdown_mic_consistent}
\\
\gamma_\uparrow
&=
R\sin^4\frac{\theta}{2}
+
r\cos^4\frac{\theta}{2}
+
\frac{\gamma_{\phi,b}}{2}\sin^2\theta ,
\label{app:gup_mic_consistent}
\\
\gamma_\phi
&=
\frac{R+r}{2}\sin^2\theta
+
\gamma_{\phi,b}\cos^2\theta .
\label{app:gphi_mic_consistent}
\end{align}
Equation~\eqref{app:Ls_mic_consistent} is precisely the same effective dressed-spin Liouvillian already introduced in Eq.~\eqref{app:Ls}; the point here is only to exhibit its microscopic origin from the bare optical recycling model \eqref{app:Ls_bare_mic_consistent}.

Using Eq.~\eqref{app:Sz0}, the stationary inversion of the uncoupled dressed spin is
\begin{equation}
S_z^{(0)}
=
\frac{\gamma_\uparrow-\gamma_\downarrow}{\gamma_\uparrow+\gamma_\downarrow}.
\label{app:Sz0_mic_start_consistent}
\end{equation}
Substituting Eqs.~\eqref{app:gdown_mic_consistent} and \eqref{app:gup_mic_consistent}, one finds
\begin{align}
\gamma_\uparrow-\gamma_\downarrow
&=
R\left(
\sin^4\frac{\theta}{2}-\cos^4\frac{\theta}{2}
\right)
+
r\left(
\cos^4\frac{\theta}{2}-\sin^4\frac{\theta}{2}
\right)
\nonumber\\
&=
-(R-r)\cos\theta ,
\label{app:Sz_num_mic_consistent}
\end{align}
while
\begin{align}
\gamma_\uparrow+\gamma_\downarrow
&=
(R+r)\left(
\sin^4\frac{\theta}{2}+\cos^4\frac{\theta}{2}
\right)
+
\gamma_{\phi,b}\sin^2\theta
\nonumber\\
&=
\frac{R+r}{2}\left(1+\cos^2\theta\right)
+
\gamma_{\phi,b}\sin^2\theta .
\label{app:Sz_den_mic_consistent}
\end{align}
Hence
\begin{equation}
S_z^{(0)}
=
-\frac{(R-r)\cos\theta}
{\dfrac{R+r}{2}\left(1+\cos^2\theta\right)+\gamma_{\phi,b}\sin^2\theta}.
\label{app:Sz0_mic_consistent}
\end{equation}
In the commonly used limit where the bare pure-dephasing contribution is neglected in the population balance, this reduces to
\begin{equation}
S_z^{(0)}
=
-\frac{2(R-r)\cos\theta}
{(R+r)\left(1+\cos^2\theta\right)}.
\label{app:Sz0_mic_simplified_consistent}
\end{equation}

Equations~\eqref{app:gdown_mic_consistent}--\eqref{app:Sz0_mic_simplified_consistent} provide the microscopic interpretation of the effective dressed rates entering Eq.~\eqref{app:Ls} used in the elimination analysis below.}

\subsection{Interaction picture and projection-operator expansion}
\label{app:proj}

We separate the dynamics into ``free'' and ``interaction'' parts. We define
\begin{equation}
H_0 = \hbar \omega_m a^\dagger a + \frac{\hbar \tilde\omega}{2}\sigma_z,\qquad
V = \hbar g (a\sigma_+ + a^\dagger \sigma_-),
\label{app:H0V}
\end{equation}
so that $H_{\mathrm{JC}}=H_0+V$. We move to the interaction picture with respect to $H_0$ \emph{and} include dissipation through $\mathcal{L}_m+\mathcal{L}_s$. Let $\rho_I(t)$ be the interaction-picture state, defined by
\begin{equation}
\rho_I(t)=e^{+iH_0 t/\hbar}\,\rho(t)\,e^{-iH_0 t/\hbar}.
\end{equation}
Then $\rho_I$ satisfies
\begin{equation}
\dot\rho_I(t)=\left(\mathcal{L}_m+\mathcal{L}_s\right)\rho_I(t)
-\frac{i}{\hbar}[V_I(t),\rho_I(t)],
\label{app:MEI}
\end{equation}
where
\begin{equation}
V_I(t)=\hbar g\Bigl(a e^{-i\omega_m t}\sigma_+ e^{+i\tilde\omega t}
+ a^\dagger e^{+i\omega_m t}\sigma_- e^{-i\tilde\omega t}\Bigr).
\label{app:VIexact}
\end{equation}
Using the detuning $\delta\equiv \tilde\omega-\omega_m$,
\begin{equation}
V_I(t)=\hbar g\Bigl(a\,\sigma_+\,e^{+i\delta t}+a^\dagger\,\sigma_-\,e^{-i\delta t}\Bigr).
\label{app:VIsimple}
\end{equation}
We now eliminate the spin using a standard second-order (Born--Markov) projection approach. We define the projection superoperator
\begin{equation}
\mathcal{P}X \equiv \Tr_s(X)\otimes \rho_s^{(0)} \equiv \rho_m \otimes \rho_s^{(0)},
\qquad \rho_m\equiv \Tr_s(X).
\label{app:projdef}
\end{equation}
Let $\mathcal{Q}=1-\mathcal{P}$. Under the assumptions in Eq.~\eqref{app:timescale}, and assuming $\mathcal{Q}\rho_I(0)=0$ (initially factorized state or rapid initial spin relaxation), the standard second-order time-local generator for $\rho_m(t)$ is (see, e.g., Nakajima--Zwanzig / time-convolutionless expansion)
\begin{equation}
\dot\rho_m(t)
= \mathcal{L}_m\rho_m(t) + \mathcal{K}\rho_m(t),
\label{app:reducedME}
\end{equation}
with
\begin{equation}\small
\mathcal{K}\rho_m
=
-\frac{1}{\hbar^2}\int_0^\infty d\tau\,
\Tr_s\!\left(
\left[V_I(t),\,e^{\mathcal{L}_s \tau}\bigl[V_I(t-\tau),\,\rho_m\otimes\rho_s^{(0)}\bigr]\right]
\right).
\label{app:Kdef}
\end{equation}
In writing Eq.~\eqref{app:Kdef} we (i) replaced the exact propagator under $\mathcal{L}_m+\mathcal{L}_s$ by $e^{\mathcal{L}_s\tau}$ inside the integral (since $\mathcal{L}_m$ is slow on the spin correlation time), and (ii) extended the upper integration limit to $\infty$ (Markov approximation) because spin correlations decay on a timescale $\sim \gamma_2^{-1}$. Equation~\eqref{app:Kdef} is the rigorous starting point for the effective mechanical Liouvillian. The remainder of the derivation consists of explicitly evaluating the double commutator and expressing the result in Lindblad form.

\subsection{Derivation of the second-order kernel $\mathcal{K}$}
\label{app:Kderivation}
In this subsection we derive Eq.~\eqref{app:Kdef} explicitly, starting from the interaction-picture master equation and using a time-convolutionless (TCL) expansion to second order in the spin--mechanics interaction. The derivation is standard but we present it in full detail to make the approximations and intermediate steps transparent.

\paragraph{Interaction-picture equation.}
Recall that in the interaction picture defined with respect to $H_0$ (and with dissipators kept explicit), the joint state $\rho_I(t)$ obeys
\begin{equation}
\dot\rho_I(t)=\bigl(\mathcal{L}_m+\mathcal{L}_s\bigr)\rho_I(t)+\mathcal{L}_I(t)\rho_I(t),
\label{app:MEI_again}
\end{equation}
where the interaction superoperator is
\begin{equation}
\mathcal{L}_I(t)X \equiv -\frac{i}{\hbar}[V_I(t),X].
\label{app:LIdef}
\end{equation}
For later use, we denote by
\begin{equation}
\mathcal{L}_0 \equiv \mathcal{L}_m+\mathcal{L}_s
\label{app:L0def}
\end{equation}
the ``free'' Liouvillian in this interaction picture (note that $H_0$ has already been removed from the commutator part).

\paragraph{Projection operators and relevant variables.}
We define the projection superoperator $\mathcal{P}$ onto the relevant subspace (mechanics) as
\begin{equation}
\mathcal{P}X \equiv \Tr_s(X)\otimes \rho_s^{(0)} \equiv \rho_m\otimes\rho_s^{(0)},\quad \rho_m\equiv \Tr_s(X),
\label{app:Pdef_again}
\end{equation}
where $\rho_s^{(0)}$ is the stationary spin state for vanishing coupling ($g=0$), satisfying
\begin{equation}
\mathcal{L}_s \rho_s^{(0)}=0.
\label{app:Ls_rhos0}
\end{equation}
The complementary projector is $\mathcal{Q}\equiv 1-\mathcal{P}$. We now list the key algebraic properties used below:
\begin{align}
\mathcal{P}^2&=\mathcal{P},\qquad \mathcal{Q}^2=\mathcal{Q},\qquad \mathcal{P}\mathcal{Q}=\mathcal{Q}\mathcal{P}=0, \label{app:proj_props}\\[2pt]
\mathcal{P}\mathcal{L}_s &= 0 \quad \text{on operators of the form } \rho_m\otimes\rho_s^{(0)}, \label{app:PLs0}\\[2pt]
\mathcal{P}\mathcal{L}_m &= \mathcal{L}_m\mathcal{P}\quad\text{(since $\mathcal{L}_m$ acts trivially on the spin).} \label{app:PLm_comm}
\end{align}
Equation~\eqref{app:PLs0} follows directly from $\mathcal{L}_s\rho_s^{(0)}=0$ and linearity of $\mathcal{L}_s$.

\paragraph{Splitting the dynamics into $\mathcal{P}$ and $\mathcal{Q}$ sectors.}
We now apply $\mathcal{P}$ and $\mathcal{Q}$ to Eq.~\eqref{app:MEI_again}. Using $\rho_I=\mathcal{P}\rho_I+\mathcal{Q}\rho_I$, we obtain two coupled equations:
\begin{align}
\frac{d}{dt}\,\mathcal{P}\rho_I(t)
&= \mathcal{P}\mathcal{L}_0\mathcal{P}\rho_I(t)
 +\mathcal{P}\mathcal{L}_0\mathcal{Q}\rho_I(t)
\nonumber\\
&\quad
 +\mathcal{P}\mathcal{L}_I(t)\mathcal{P}\rho_I(t)
 +\mathcal{P}\mathcal{L}_I(t)\mathcal{Q}\rho_I(t),
\label{app:P_eq_full}\\[4pt]
\frac{d}{dt}\,\mathcal{Q}\rho_I(t)
&= \mathcal{Q}\mathcal{L}_0\mathcal{P}\rho_I(t)
 +\mathcal{Q}\mathcal{L}_0\mathcal{Q}\rho_I(t)
\nonumber\\
&\quad
 +\mathcal{Q}\mathcal{L}_I(t)\mathcal{P}\rho_I(t)
 +\mathcal{Q}\mathcal{L}_I(t)\mathcal{Q}\rho_I(t).
\label{app:Q_eq_full}
\end{align}
We now simplify these equations using the properties above. First,
\begin{equation}
\mathcal{P}\mathcal{L}_0\mathcal{P}=\mathcal{P}(\mathcal{L}_m+\mathcal{L}_s)\mathcal{P}
=\mathcal{P}\mathcal{L}_m\mathcal{P}+\underbrace{\mathcal{P}\mathcal{L}_s\mathcal{P}}_{=\,0}
=\mathcal{L}_m\mathcal{P}.
\label{app:PL0P}
\end{equation}
Second, because $\mathcal{L}_m$ acts only on the mechanics and $\mathcal{P}$ fixes the spin to $\rho_s^{(0)}$, we have
\begin{equation}
\mathcal{Q}\mathcal{L}_m\mathcal{P}=0,
\label{app:QLmP}
\end{equation}
since $\mathcal{L}_m\mathcal{P}\rho_I=\mathcal{L}_m(\rho_m\otimes\rho_s^{(0)})$ still lies entirely in the range of $\mathcal{P}$. Furthermore,
\begin{equation}
\mathcal{Q}\mathcal{L}_s\mathcal{P}=0
\label{app:QLsP}
\end{equation}
because $\mathcal{L}_s(\rho_m\otimes\rho_s^{(0)})=\rho_m\otimes(\mathcal{L}_s\rho_s^{(0)})=0$.
Therefore $\mathcal{Q}\mathcal{L}_0\mathcal{P}=0$. With these simplifications, Eqs.~\eqref{app:P_eq_full}--\eqref{app:Q_eq_full} become
\begin{align}
\frac{d}{dt}\,\mathcal{P}\rho_I(t)
&= \mathcal{L}_m\,\mathcal{P}\rho_I(t)
+\mathcal{P}\mathcal{L}_0\mathcal{Q}\rho_I(t)
\nonumber\\
&\quad
+\mathcal{P}\mathcal{L}_I(t)\mathcal{P}\rho_I(t)
+\mathcal{P}\mathcal{L}_I(t)\mathcal{Q}\rho_I(t),
\label{app:P_eq_simpl}\\[4pt]
\frac{d}{dt}\,\mathcal{Q}\rho_I(t)
&= \mathcal{Q}\mathcal{L}_0\mathcal{Q}\rho_I(t)
+\mathcal{Q}\mathcal{L}_I(t)\mathcal{P}\rho_I(t)
\nonumber\\
&\quad
+\mathcal{Q}\mathcal{L}_I(t)\mathcal{Q}\rho_I(t).
\label{app:Q_eq_simpl}
\end{align}

\paragraph{Formal solution for $\mathcal{Q}\rho_I(t)$.}
Equation~\eqref{app:Q_eq_simpl} is an inhomogeneous linear equation for $\mathcal{Q}\rho_I(t)$.
We define the propagator in the $\mathcal{Q}$ subspace generated by $\mathcal{Q}\mathcal{L}_0\mathcal{Q}$,
\begin{equation}
\mathcal{G}(t,t_0) \equiv \mathcal{T}\exp\!\left[\int_{t_0}^{t} ds\,\mathcal{Q}\mathcal{L}_0\mathcal{Q}\right],
\label{app:Gdef}
\end{equation}
where $\mathcal{T}$ denotes time ordering. Since $\mathcal{L}_0$ is time-independent here, this reduces to $\mathcal{G}(t,t_0)=e^{\mathcal{Q}\mathcal{L}_0\mathcal{Q}(t-t_0)}$. Treating the interaction as a perturbation and retaining terms up to second order in $\mathcal{L}_I$, we neglect the term $\mathcal{Q}\mathcal{L}_I(t)\mathcal{Q}\rho_I(t)$ inside Eq.~\eqref{app:Q_eq_simpl} when solving for $\mathcal{Q}\rho_I(t)$ (this is the standard Born approximation consistent with truncation at $\mathcal{O}(g^2)$). We then have
\begin{equation}
\frac{d}{dt}\,\mathcal{Q}\rho_I(t)
\simeq \mathcal{Q}\mathcal{L}_0\mathcal{Q}\rho_I(t)
+\mathcal{Q}\mathcal{L}_I(t)\mathcal{P}\rho_I(t).
\label{app:Qeq_Born}
\end{equation}
The formal solution is
\begin{equation}
\mathcal{Q}\rho_I(t)\simeq \mathcal{G}(t,0)\,\mathcal{Q}\rho_I(0)
+\int_0^t ds\,\mathcal{G}(t,s)\,\mathcal{Q}\mathcal{L}_I(s)\mathcal{P}\rho_I(s).
\label{app:Qsol}
\end{equation}
We now assume
\begin{equation}
\mathcal{Q}\rho_I(0)=0,
\label{app:Qinit0}
\end{equation}
which holds for an initially factorized state $\rho_I(0)=\rho_m(0)\otimes\rho_s^{(0)}$ and is also justified if the spin relaxes rapidly to $\rho_s^{(0)}$ on timescales short compared to the mechanical evolution. Under Eq.~\eqref{app:Qinit0},
\begin{equation}
\mathcal{Q}\rho_I(t)\simeq \int_0^t ds\,\mathcal{G}(t,s)\,\mathcal{Q}\mathcal{L}_I(s)\mathcal{P}\rho_I(s).
\label{app:Qsol2}
\end{equation}

\paragraph{Closed equation for $\mathcal{P}\rho_I(t)$ to second order.}
Let us insert Eq.~\eqref{app:Qsol2} into Eq.~\eqref{app:P_eq_simpl}. Keeping terms up to $\mathcal{O}(g^2)$, we may (i) drop the term $\mathcal{P}\mathcal{L}_0\mathcal{Q}\rho_I$ because it is already $\mathcal{O}(g)$ and would multiply $\mathcal{Q}\rho_I=\mathcal{O}(g)$ only through higher-order feedback, and (ii) drop $\mathcal{P}\mathcal{L}_I(t)\mathcal{Q}\rho_I(t)$ beyond substituting $\mathcal{Q}\rho_I$ at leading order from Eq.~\eqref{app:Qsol2}. We obtain the standard second-order TCL generator:
\begin{align}
\frac{d}{dt}\,\mathcal{P}\rho_I(t)
&\simeq \mathcal{L}_m\,\mathcal{P}\rho_I(t)
+\mathcal{P}\mathcal{L}_I(t)\mathcal{P}\rho_I(t)
\nonumber\\
&\quad
+\int_0^t ds\,\mathcal{P}\mathcal{L}_I(t)\,\mathcal{G}(t,s)\,\mathcal{Q}\mathcal{L}_I(s)\,\mathcal{P}\rho_I(s).
\label{app:TCL2_preMarkov}
\end{align}

\paragraph{First-order term $\mathcal{P}\mathcal{L}_I(t)\mathcal{P}$.}
For the present interaction $V_I(t)\propto a\sigma_+ + a^\dagger\sigma_-$ and a stationary spin state $\rho_s^{(0)}$ diagonal in the dressed basis, we have $\Tr_s(\sigma_\pm\rho_s^{(0)})=0$. Hence
\begin{equation}
\mathcal{P}\mathcal{L}_I(t)\mathcal{P}\rho_I(t)
=
-\frac{i}{\hbar}\Tr_s\!\left([V_I(t),\rho_m(t)\otimes\rho_s^{(0)}]\right)\otimes\rho_s^{(0)}
=0.
\label{app:PLIP_zero}
\end{equation}
Thus the leading nontrivial contribution is second order.

\paragraph{Markov approximation and reduction to Eq.~\eqref{app:Kdef}.}
We now exploit the separation of timescales $\gamma_2^{-1}\ll$ (mechanical evolution time). The integrand in Eq.~\eqref{app:TCL2_preMarkov} contains the propagator $\mathcal{G}(t,s)\approx e^{\mathcal{Q}\mathcal{L}_0\mathcal{Q}(t-s)}$ and decays on the spin correlation time $\sim \gamma_2^{-1}$. Therefore we may:
\begin{enumerate}
\item Replace $\rho_I(s)$ by $\rho_I(t)$ inside the integral (time-local approximation),
\item Extend the upper limit $t\to\infty$ (Markov limit),
\item Replace $\mathcal{G}(t,s)$ by $e^{\mathcal{L}_s (t-s)}$ inside the trace over the spin, since $\mathcal{L}_m$ is slow on the spin correlation time.
\end{enumerate}
Implementing these steps and writing $\tau=t-s$ yields
\begin{align}
\frac{d}{dt}\,\mathcal{P}\rho_I(t)
&\simeq \mathcal{L}_m\,\mathcal{P}\rho_I(t)
+\int_0^\infty d\tau\,
\mathcal{P}\mathcal{L}_I(t)\,e^{\mathcal{L}_s \tau}\,\mathcal{L}_I(t-\tau)\,\mathcal{P}\rho_I(t).
\label{app:TCL2_Markov}
\end{align}
Finally, using $\mathcal{L}_I X=-(i/\hbar)[V_I,X]$ and $\mathcal{P}\rho_I(t)=\rho_m(t)\otimes\rho_s^{(0)}$, we obtain
\begin{equation}
\dot\rho_m(t) = \mathcal{L}_m\rho_m(t) + \mathcal{K}\rho_m(t),
\end{equation}
with the second-order kernel
\begin{equation}
\mathcal{K}\rho_m
=
-\frac{1}{\hbar^2}\int_0^\infty d\tau\,
\Tr_s\!\left(
\left[V_I(t),\,e^{\mathcal{L}_s \tau}\bigl[V_I(t-\tau),\,\rho_m\otimes\rho_s^{(0)}\bigr]\right]
\right),
\end{equation}
which is Eq.~\eqref{app:Kdef}.

\paragraph{Comments on the approximations.}
The derivation above makes explicit that Eq.~\eqref{app:Kdef} relies on: (i) a perturbative truncation at $\mathcal{O}(g^2)$ (Born approximation), (ii) rapid decay of spin correlations (Markov approximation), and (iii) separation of timescales that allows one to treat the mechanical state as quasi-static over the spin correlation time. These conditions are summarized in Eq.~\eqref{app:timescale}.

\subsection{Explicit evaluation of the second-order kernel: from the double commutator to Lindblad form}
\label{app:kernel_improved}

In this subsection we evaluate the kernel \eqref{app:Kdef} explicitly up to the point where it is expressed in terms of stationary spin correlation functions and written in a manifest Lindblad $+$ Lamb-shift form.
No explicit form of the correlators is assumed here; they are computed in the next subsection. 
\paragraph{Step 1: convenient decomposition of $V_I(t)$.}
Using Eq.~\eqref{app:VIsimple}, we write
\begin{equation}
V_I(t)=\hbar\Bigl(A\,e^{+i\delta t}+A^\dagger e^{-i\delta t}\Bigr),
\qquad
A\equiv g\,a\,\sigma_+,
\label{app:AI_improved}
\end{equation}
so that $A^\dagger=g\,a^\dagger\sigma_-$. In terms of the interaction superoperator
$\mathcal{L}_I(t)X\equiv-\frac{i}{\hbar}[V_I(t),X]$ the second-order generator can be written as
\begin{equation}
\mathcal{K}\rho_m
=
\int_0^\infty d\tau\;
\Tr_s\!\left\{
\mathcal{L}_I(t)\,e^{\mathcal{L}_s\tau}\,\mathcal{L}_I(t-\tau)
\bigl(\rho_m\otimes\rho_s^{(0)}\bigr)
\right\},
\label{app:K_superop}
\end{equation}
which is equivalent to Eq.~\eqref{app:Kdef}.

\paragraph{Step 2: expand the inner commutator.}
Let $X\equiv\rho_m\otimes\rho_s^{(0)}$ and define
\begin{equation}
Y(\tau)\equiv e^{\mathcal{L}_s\tau}\bigl[V_I(t-\tau),X\bigr].
\label{app:Ydef}
\end{equation}
Then Eq.~\eqref{app:Kdef} reads
\begin{equation}
\mathcal{K}\rho_m=-\frac{1}{\hbar^2}\int_0^\infty d\tau\;\Tr_s\!\bigl([V_I(t),Y(\tau)]\bigr).
\label{app:Kdef_again}
\end{equation}
From \eqref{app:AI_improved},
\begin{align}
[V_I(t-\tau),X]
&=\hbar e^{+i\delta(t-\tau)}\,[A,X]+\hbar e^{-i\delta(t-\tau)}\,[A^\dagger,X].
\label{app:inner_split}
\end{align}
Using that $a$ commutes with all spin operators,
\begin{align}
[A,X]
&=g\Bigl(a\rho_m\otimes\sigma_+\rho_s^{(0)}-\rho_m a\otimes\rho_s^{(0)}\sigma_+\Bigr),
\label{app:Acomm_explicit}\\
[A^\dagger,X]
&=g\Bigl(a^\dagger\rho_m\otimes\sigma_-\rho_s^{(0)}-\rho_m a^\dagger\otimes\rho_s^{(0)}\sigma_-\Bigr).
\label{app:Adagcomm_explicit}
\end{align}
Thus the only spin operators that ever enter the propagator $e^{\mathcal{L}_s\tau}$ are $\sigma_\pm\rho_s^{(0)}$ and $\rho_s^{(0)}\sigma_\pm$.

\paragraph{Step 3: apply $e^{\mathcal{L}_s\tau}$ and identify the surviving spin structures.}

Since $\mathcal{L}_s$ acts only on the spin Hilbert space, we have for any mechanical operator $M$, and any spin operator $S$
\begin{equation}
e^{\mathcal{L}_s\tau}(M\otimes S)=M\otimes e^{\mathcal{L}_s\tau}(S).
\label{app:Ls_factor}
\end{equation}
Applying $e^{\mathcal{L}_s\tau}$ to \eqref{app:inner_split}--\eqref{app:Adagcomm_explicit} gives
\begin{align}
Y(\tau)
&=\hbar g\,e^{+i\delta(t-\tau)}
\Bigl[
\begin{aligned}[t]
& a\rho_m\otimes e^{\mathcal{L}_s\tau}(\sigma_+\rho_s^{(0)})
\\
&-\rho_m a\otimes e^{\mathcal{L}_s\tau}(\rho_s^{(0)}\sigma_+)
\end{aligned}
\Bigr]
\nonumber\\
&\quad+\hbar g\,e^{-i\delta(t-\tau)}
\Bigl[
\begin{aligned}[t]
& a^\dagger\rho_m\otimes e^{\mathcal{L}_s\tau}(\sigma_-\rho_s^{(0)})
\\
&-\rho_m a^\dagger\otimes e^{\mathcal{L}_s\tau}(\rho_s^{(0)}\sigma_-)
\end{aligned}
\Bigr].
\label{app:Y_explicit}
\end{align}
We now insert \eqref{app:Y_explicit} into the outer commutator $[V_I(t),Y(\tau)]$. Because $V_I(t)$ contains only $A\propto\sigma_+$ and $A^\dagger\propto\sigma_-$, the outer commutator produces terms involving products $\sigma_+\cdots\sigma_+$, $\sigma_-\cdots\sigma_-$, and cross terms $\sigma_+\cdots\sigma_-$ or $\sigma_-\cdots\sigma_+$. The ``equal'' products vanish identically because $\sigma_+^2=\sigma_-^2=0$. Therefore, only the cross terms survive, and these are precisely the ones that generate the stationary correlators. Concretely, one of the surviving contributions comes from the outer $A^\dagger e^{-i\delta t}$ acting on the inner $A e^{+i\delta(t-\tau)}$ part of $Y(\tau)$; the accumulated phase is $e^{-i\delta t}\,e^{+i\delta(t-\tau)}=e^{-i\delta\tau}$ and the spin trace produces 
\begin{equation}
C_{-+}(\tau)\equiv
\Tr_s\!\left[
\sigma_-\,e^{\mathcal{L}_s\tau}\!\bigl(\sigma_+\rho_s^{(0)}\bigr)
\right].
\label{app:Cmp_def}
\end{equation}
Similarly, the outer $A e^{+i\delta t}$ acting on the inner $A^\dagger e^{-i\delta(t-\tau)}$ part produces the phase $e^{+i\delta\tau}$ and the correlator
\begin{equation}
C_{+-}(\tau)\equiv
\Tr_s\!\left[
\sigma_+\,e^{\mathcal{L}_s\tau}\!\bigl(\sigma_-\rho_s^{(0)}\bigr)
\right].
\label{app:Cpm_def}
\end{equation}
These coincide with the standard stationary two-time correlators, $C_{-+}(\tau)=\langle\sigma_-(\tau)\sigma_+(0)\rangle_{\rm ss}$ and $C_{+-}(\tau)=\langle\sigma_+(\tau)\sigma_-(0)\rangle_{\rm ss}$, via $\langle X(\tau)Y(0)\rangle_{\rm ss}=\Tr_s\!\left[X\,e^{\mathcal{L}_s\tau}(Y\rho_s^{(0)})\right]$.

\paragraph{Step 4: explicit kernel after tracing over the spin.}
Performing the commutator algebra and retaining only the cross terms that survive the spin trace yields
\begin{align}
\mathcal{K}\rho_m
&=
g^2\int_0^\infty d\tau\,
\Bigl\{
e^{-i\delta\tau}\,C_{-+}(\tau)\,\Bigl(a\,\rho_m a^\dagger - a^\dagger a\,\rho_m\Bigr)
\nonumber\\
&\hspace{1.2cm}
+e^{+i\delta\tau}\,C_{+-}(\tau)\,\Bigl(a^\dagger\,\rho_m a - a a^\dagger\,\rho_m\Bigr)
+\mathrm{H.c.}
\Bigr\}.
\label{app:Kexpanded_improved}
\end{align}
The Hermitian-conjugate part collects the terms where the operator order in the outer commutator is reversed and guarantees that $\mathcal{K}$ maps Hermitian operators to Hermitian operators.

\paragraph{Step 5: one-sided Fourier transforms and separation into dissipative and Hamiltonian parts.}
We define the one-sided Fourier transforms (susceptibilities)
\begin{equation}
\begin{aligned}
\Xi_{-+}(\delta) &\equiv \int_{0}^{\infty}\! d\tau\, e^{-i\delta\tau}\, C_{-+}(\tau),\\
\Xi_{+-}(\delta) &\equiv \int_{0}^{\infty}\! d\tau\, e^{+i\delta\tau}\, C_{+-}(\tau).
\end{aligned}
\label{app:Xidef_improved}
\end{equation}
Then Eq.~\eqref{app:Kexpanded_improved} becomes
\begin{align}
\mathcal{K}\rho_m
&=
g^2\Big[
\Xi_{-+}(\delta)\Bigl(a\rho_m a^\dagger-a^\dagger a\rho_m\Bigr)
+\Xi_{+-}(\delta)\Bigl(a^\dagger\rho_m a-a a^\dagger\rho_m\Bigr)
\nonumber\\
&
+\Xi_{-+}^\ast(\delta)\Bigl(a\rho_m a^\dagger-\rho_m a^\dagger a\Bigr)
+\Xi_{+-}^\ast(\delta)\Bigl(a^\dagger\rho_m a-\rho_m a a^\dagger\Bigr)
\Big].
\label{app:KXi_improved}
\end{align}
To identify the dissipative and Hamiltonian contributions, we use the operator identities
\begin{align}
\Bigl(a\rho a^\dagger-a^\dagger a\rho\Bigr)+\Bigl(a\rho a^\dagger-\rho a^\dagger a\Bigr)
&= 2\mathcal{D}[a]\rho,
\label{app:id_Da}\\
\Bigl(a^\dagger\rho a-a a^\dagger\rho\Bigr)+\Bigl(a^\dagger\rho a-\rho a a^\dagger\Bigr)
&= 2\mathcal{D}[a^\dagger]\rho,
\label{app:id_Dadag}
\end{align}
and
\begin{align}
\Bigl(a\rho a^\dagger-a^\dagger a\rho\Bigr)-\Bigl(a\rho a^\dagger-\rho a^\dagger a\Bigr)
&= [\rho,a^\dagger a],
\label{app:id_comm1}\\
\Bigl(a^\dagger\rho a-a a^\dagger\rho\Bigr)-\Bigl(a^\dagger\rho a-\rho a a^\dagger\Bigr)
&= [\rho,a a^\dagger]=[\rho,a^\dagger a],
\label{app:id_comm2}
\end{align}
where we used $a a^\dagger=a^\dagger a+\mathbf{1}$ and $[\rho,\mathbf{1}]=0$. 
Using the identities \eqref{app:id_Da}--\eqref{app:id_comm2}, we separate Eq.~\eqref{app:KXi_improved} into dissipative and Hamiltonian contributions:
\begin{equation}
\mathcal{K}\rho_m
= \Gamma_-\,\mathcal{D}[a]\rho_m + \Gamma_+\,\mathcal{D}[a^\dagger]\rho_m
-\frac{i}{\hbar}\,[H_{\mathrm{LS}},\rho_m],
\label{app:Kfinal_improved}
\end{equation}
where the transition rates are
\begin{equation}
\Gamma_- = 2g^2\,\Re\bigl[\Xi_{-+}(\delta)\bigr],\qquad
\Gamma_+ = 2g^2\,\Re\bigl[\Xi_{+-}(\delta)\bigr],
\label{app:GammasXi_improved}
\end{equation}
and the Lamb-shift Hamiltonian is
\begin{equation}
H_{\mathrm{LS}} = \hbar\,\delta\omega\, a^\dagger a,\qquad
\delta\omega = -g^2\Big(\Im\bigl[\Xi_{-+}(\delta)\bigr]+\Im\bigl[\Xi_{+-}(\delta)\bigr]\Big).
\label{app:Lamb_improved}
\end{equation}
Equations~\eqref{app:Kfinal_improved}--\eqref{app:Lamb_improved} are general within the controlled second-order elimination. The remaining task is to compute $C_{-+}(\tau)$ and $C_{+-}(\tau)$ for the specific spin Liouvillian \eqref{app:Ls}, which is done in the next subsection.

\subsection{Spin correlation functions and evaluation of $\Gamma_\pm$}
\label{app:QRT_rates_clean}
The stationary correlators entering Eqs.~\eqref{app:Cmp_def}--\eqref{app:Cpm_def} are
\begin{equation}
C_{-+}(\tau)\equiv \langle \sigma_-(\tau)\sigma_+(0)\rangle_{\rm ss}
= \Tr_s\!\left[\sigma_-\,e^{\mathcal{L}_s\tau}\bigl(\sigma_+ \rho_s^{(0)}\bigr)\right],
\label{app:Cmpss}
\end{equation}
\begin{equation}
C_{+-}(\tau)\equiv \langle \sigma_+(\tau)\sigma_-(0)\rangle_{\rm ss}
= \Tr_s\!\left[\sigma_+\,e^{\mathcal{L}_s\tau}\bigl(\sigma_- \rho_s^{(0)}\bigr)\right],
\label{app:Cpmss}
\end{equation}
where $\tau\ge 0$ and $\rho_s^{(0)}$ is the stationary spin state for $g=0$, given in Eq.~\eqref{app:rhos0}. From $\sigma_-\sigma_+=\ket{\downarrow}\!\bra{\downarrow}$ and $\sigma_+\sigma_-=\ket{\uparrow}\!\bra{\uparrow}$ we immediately have
\begin{equation}
C_{-+}(0)=\Tr_s(\sigma_-\sigma_+\rho_s^{(0)})=p_\downarrow ,
\label{app:C0vals1}
\end{equation}
and
\begin{equation}
C_{+-}(0)=\Tr_s(\sigma_+\sigma_-\rho_s^{(0)})=p_\uparrow.
\label{app:C0vals2}
\end{equation}
To obtain $C_{\mp\pm}(\tau)$ we first determine the decay of $\sigma_\pm(\tau)$ under the adjoint Liouvillian $\mathcal{L}_s^\dagger$. Using the identity
\begin{equation}
\mathcal{D}^\dagger[L]O = L^\dagger O L - \frac12\{L^\dagger L,O\},
\end{equation}
together with the spin Liouvillian \eqref{app:Ls}, one finds
\begin{equation}
\begin{aligned}
\mathcal{D}^\dagger[\sigma_-]\sigma_- &= -\frac12\,\sigma_-,\\
\mathcal{D}^\dagger[\sigma_+]\sigma_- &= -\frac12\,\sigma_-,\\
\mathcal{D}^\dagger[\sigma_z]\sigma_- &= -2\,\sigma_-.
\end{aligned}
\end{equation}
and therefore
\begin{equation}
\begin{aligned}
\frac{d}{d\tau}\sigma_-(\tau)
&= \mathcal{L}_s^\dagger \sigma_-(\tau)\\
&= -\left(\frac{\gamma_\uparrow+\gamma_\downarrow}{2}+\gamma_\phi\right)\sigma_-(\tau)\\
&\equiv -\gamma_2\,\sigma_-(\tau).
\end{aligned}
\label{app:sigmaminus_decay}
\end{equation}
with $\gamma_2$ defined in Eq.~\eqref{app:gamma12}. The same calculation gives
\begin{equation}
\frac{d}{d\tau}\sigma_+(\tau) = -\gamma_2\,\sigma_+(\tau).
\label{app:sigmaplus_decay}
\end{equation}
By the quantum regression theorem, the two-time correlators \eqref{app:Cmpss}--\eqref{app:Cpmss} obey the same homogeneous equations with initial conditions fixed by \eqref{app:C0vals1} and \eqref{app:C0vals2}. Hence
\begin{equation}
\begin{aligned}
\frac{d}{d\tau}C_{-+}(\tau) &= -\gamma_2\, C_{-+}(\tau), \qquad C_{-+}(0)=p_\downarrow,\\
\Longrightarrow\qquad
C_{-+}(\tau) &= p_\downarrow\, e^{-\gamma_2\tau}.
\end{aligned}
\label{app:Cmp_clean}
\end{equation}
As well we have
\begin{equation}
\begin{aligned}
\frac{d}{d\tau}C_{+-}(\tau) &= -\gamma_2\, C_{+-}(\tau), \qquad C_{+-}(0)=p_\uparrow,\\
\Longrightarrow\qquad
C_{+-}(\tau) &= p_\uparrow\, e^{-\gamma_2\tau}.
\end{aligned}
\label{app:Cpm_clean}
\end{equation}
We now evaluate the susceptibilities (one-sided Fourier transforms) appearing in the previous subsection, with the phase factors fixed by the detuning $\delta$,
\begin{equation}
\begin{aligned}
\Xi_{-+}(\delta)
&\equiv \int_0^\infty d\tau\, e^{+i\delta\tau}\,C_{-+}(\tau)\\
&= \int_0^\infty d\tau\, e^{-(\gamma_2-i\delta)\tau}\\
&= \frac{p_\downarrow}{\gamma_2-i\delta}.
\end{aligned}
\label{app:Xi_m}
\end{equation}
As well we have
\begin{equation}
\begin{aligned}
\Xi_{+-}(\delta)
&\equiv \int_0^\infty d\tau\, e^{-i\delta\tau}\,C_{+-}(\tau)\\
&= \int_0^\infty d\tau\, e^{-(\gamma_2+i\delta)\tau}\\
&= \frac{p_\uparrow}{\gamma_2+i\delta}.
\end{aligned}
\label{app:Xi_p}
\end{equation}
Using the general relations obtained from the kernel decomposition (cf.\ Eq.~\eqref{app:Kfinal_improved}),
\begin{equation}
\Gamma_- = 2g^2\,\Re\!\left[\Xi_{-+}(\delta)\right],
\qquad
\Gamma_+ = 2g^2\,\Re\!\left[\Xi_{+-}(\delta)\right],
\label{app:GammasXi_clean}
\end{equation}
we arrive at
\begin{equation}
\Gamma_-(\delta)=2g^2\,p_\downarrow \,\frac{\gamma_2}{\gamma_2^2+\delta^2},
\qquad
\Gamma_+(\delta)=2g^2\,p_\uparrow\,\frac{\gamma_2}{\gamma_2^2+\delta^2}.
\label{app:Gammas_Lorentz}
\end{equation}
On resonance ($\delta=0$),
\begin{equation}
\Gamma_-=\frac{2g^2}{\gamma_2}\,p_\downarrow ,\qquad
\Gamma_+=\frac{2g^2}{\gamma_2}\,p_\uparrow.
\label{app:Gammas_res}
\end{equation}
The spin-induced contribution to the \emph{phonon-number} damping at zero detuning is therefore
\begin{equation}
\Gamma_{\mathrm{NV}}\equiv \Gamma_- - \Gamma_+
= \frac{2g^2}{\gamma_2}(p_\downarrow -p_\uparrow)
= -\frac{2g^2}{\gamma_2}S_z^{(0)},
\label{app:Gamma_opt_clean}
\end{equation}
where $S_z^{(0)}=p_\uparrow-p_\downarrow $ defined in Eq.~\eqref{app:Sz0}. Including intrinsic mechanical damping $\gamma_m$, the net phonon-number damping reads
\begin{equation}
\gamma_{\rm eff}^{(n)}=\gamma_m+\Gamma_{\mathrm{NV}}
=\gamma_m-\frac{2g^2}{\gamma_2}S_z^{(0)}.
\label{app:gamma_eff_clean}
\end{equation}
Thus inversion ($S_z^{(0)}>0$) reduces damping and yields gain when $S_z^{(0)}>\gamma_m\gamma_2/(2g^2)$.

\subsection{Closed-form gain and damping rates, threshold condition, and sign consistency}
\label{app:rates}

The elimination procedure yields a reduced mechanical master equation consisting of the intrinsic mechanical bath terms \eqref{app:Lm} supplemented by the spin-induced kernel \eqref{app:Kfinal_improved}. The latter contributes the two additional jump processes $\Gamma_-\,\mathcal{D}[a]$ and $\Gamma_+\,\mathcal{D}[a^\dagger]$, with $\Gamma_\pm(\delta)$ obtained explicitly in Eq.~\eqref{app:Gammas_Lorentz}. It is useful to collect these contributions into the total upward and downward transition rates of the mechanical mode,
\begin{equation}
A_- \equiv \gamma_m(\bar n_{\rm th}+1)+\Gamma_-,
\qquad
A_+ \equiv \gamma_m\bar n_{\rm th}+\Gamma_+.
\label{app:Apm_def}
\end{equation}
With this notation, the reduced dynamics takes the standard form $\dot\rho_m=\cdots + A_-\,\mathcal{D}[a]\rho_m + A_+\,\mathcal{D}[a^\dagger]\rho_m$ (the ellipsis denotes the Hamiltonian part, including the Lamb shift). A direct and transparent way to identify the net damping (or gain) is to write the equation of motion for the mean phonon number $\bar n(t)\equiv\langle a^\dagger a\rangle$. Using the Lindblad identities for $\mathcal{D}[a]$ and $\mathcal{D}[a^\dagger]$, one finds
\begin{equation}
\frac{d}{dt}\bar n(t)=-(A_- - A_+)\,\bar n(t)+A_+.
\label{app:ndot_general}
\end{equation}
The coefficient of $\bar n(t)$ defines the effective \emph{phonon-number} damping rate,
\begin{equation}
\gamma_{\rm eff}^{(n)}(\delta)=A_- - A_+ = \gamma_m+\Gamma_- (\delta)-\Gamma_+(\delta)
\equiv \gamma_m+\Gamma_{\mathrm{NV}}(\delta),
\label{app:gamma_eff_from_Apm}
\end{equation}
which is consistent with the definition of $\Gamma_{\mathrm{NV}}$ given in Eq.~\eqref{app:Gamma_opt_clean}. In the stable regime $\gamma_{\rm eff}^{(n)}(\delta)>0$ (below threshold), Eq.~\eqref{app:ndot_general} relaxes to the steady-state occupation
\begin{equation}
\bar n_{\rm ss}(\delta)=\frac{A_+}{A_- - A_+}
=\frac{\gamma_m\bar n_{\rm th}+\Gamma_+(\delta)}{\gamma_m+\Gamma_- (\delta)-\Gamma_+(\delta)}.
\label{app:nss}
\end{equation}
Equation~\eqref{app:nss} makes explicit that the spin not only modifies the damping through $\Gamma_- - \Gamma_+$ but also contributes additional diffusion through $\Gamma_+$. The lasing (maser) threshold corresponds to the point where the linear damping vanishes, $\gamma_{\rm eff}^{(n)}(0)=0$, at which $\bar n_{\rm ss}$ diverges and the linear theory breaks down. Using Eq.~\eqref{app:Gamma_opt_clean} at $\delta=0$ yields the threshold inversion
\begin{equation}
S_z^{(\mathrm{th})}=\frac{\gamma_m\gamma_2}{2g^2},
\label{app:Szth}
\end{equation}
and more generally, keeping the detuning dependence in Eq.~\eqref{app:Gammas_Lorentz}, the threshold condition can be written as
\begin{equation}
S_z^{(\mathrm{th})}(\delta)=\frac{\gamma_m}{2g^2}\,\frac{\gamma_2^2+\delta^2}{\gamma_2}.
\label{app:Szth_detuned}
\end{equation}
These expressions emphasize the sign structure: inversion $S_z^{(0)}>0$ implies $\Gamma_{\mathrm{NV}}(\delta)<0$ Eq.~\eqref{app:Gamma_opt_clean}, so the spin reduces the net damping and can render it negative, leading to gain. Finally, the coherent frequency renormalization (Lamb shift) in Eq.~\eqref{app:Kfinal_improved} follows from Eq.~\eqref{app:Lamb_improved}. Substituting the susceptibilities \eqref{app:Xi_m}--\eqref{app:Xi_p} gives an odd-in-detuning shift of the mechanical frequency,
\begin{equation}
\begin{aligned}
\delta\omega(\delta)
&= -g^2\!\left[\Im\Xi_{-+}(\delta)+\Im\Xi_{+-}(\delta)\right]\\
&= -g^2\,S_z^{(0)}\,\frac{\delta}{\gamma_2^2+\delta^2}.
\end{aligned}
\label{app:freqshift}
\end{equation}
which is small in the weak-coupling regime and consistent with linear response.

\emph{Remark (sign consistency).}
Equations~\eqref{app:gamma_eff_from_Apm}--\eqref{app:Szth_detuned} make explicit that positive inversion ($S_z^{(0)}>0$) reduces the effective damping and can make it negative.

\subsection{Semiclassical gain saturation and steady-state coherent phonon number}
\label{app:sat}

The linear (small-signal) analysis determines the gain coefficient and the threshold condition. Above the threshold, the mechanical field grows until the dressed spin inversion is reduced by stimulated emission into the oscillator, resulting in gain clamping. Here we derive the steady-state coherent phonon number in a Maxwell--Bloch (mean-field) treatment, keeping the steps explicit and using the same interaction picture. We specialize to exact resonance, $\delta=\tilde\omega-\omega_m=0$, so that all operators are slowly varying (the free precession has been removed). We introduce the mean fields
\begin{equation}
\alpha \equiv \langle a\rangle,\qquad s\equiv \langle \sigma_-\rangle,\qquad
w \equiv \langle \sigma_z\rangle .
\label{app:defs_mf}
\end{equation}
Their equations of motion follow from the adjoint master equation, $\frac{d}{dt}\langle O\rangle=\Tr\!\left[(\mathcal{L}_m^\dagger+\mathcal{L}_s^\dagger)\,O\,\rho\right] -\frac{i}{\hbar}\langle[O,H_{\rm JC}]\rangle$, using the same $\mathcal{L}_m$ and $\mathcal{L}_s$ as in Eqs.~\eqref{app:Lm} and \eqref{app:Ls}. In the present interaction picture and at $\delta=0$, the coherent coupling is generated by the resonant term $\hbar g(a\sigma_+ + a^\dagger\sigma_-)$, yielding
\begin{align}
\dot\alpha &= -\frac{\gamma_m}{2}\alpha - i g\,s,
\label{app:eq_alpha}\\
\dot s &= -\gamma_2\,s + i g\,\langle a\sigma_z\rangle,
\label{app:eq_s_exact}\\
\dot w &= -\gamma_1\left(w - S_z^{(0)}\right) + 2 i g\left(\langle a^\dagger\sigma_-\rangle-\langle a\sigma_+\rangle\right).
\label{app:eq_w_exact}
\end{align}
To close the system we employ the standard semiclassical factorization,
\begin{equation}
\langle a\sigma_z\rangle \approx \langle a\rangle\langle\sigma_z\rangle=\alpha w,
\quad
\langle a^\dagger\sigma_-\rangle \approx \alpha^\ast s,
\quad
\langle a\sigma_+\rangle \approx \alpha s^\ast,
\label{app:mf_factorization}
\end{equation}
which becomes accurate once the mechanical field is macroscopic. With Eq.~\eqref{app:mf_factorization}, Eqs.~\eqref{app:eq_s_exact}--\eqref{app:eq_w_exact} reduce to
\begin{align}
\dot s &= -\gamma_2\,s + i g\,\alpha\,w,
\label{app:eq_s}\\
\dot w &= -\gamma_1\left(w - S_z^{(0)}\right) + 2 i g\left(\alpha^\ast s - \alpha s^\ast\right),
\label{app:eq_w}
\end{align}
together with Eq.~\eqref{app:eq_alpha}. We now solve the steady state $\dot\alpha=\dot s=\dot w=0$. The equations are invariant under a global phase rotation $\alpha\to \alpha e^{i\varphi}$ and $s\to s e^{i\varphi}$, hence we may fix the steady-state phase such that
\begin{equation}
\alpha=\alpha^\ast \ge 0.
\label{app:alphareal}
\end{equation}
\paragraph{Step 1: eliminate $s$.}
From $\dot s=0$ in Eq.~\eqref{app:eq_s},
\begin{equation}
s = \frac{i g}{\gamma_2}\,\alpha\,w.
\label{app:s_solution}
\end{equation}

\paragraph{Step 2: inversion depletion $w(\alpha)$.}
Using Eq.~\eqref{app:s_solution} and Eq.~\eqref{app:alphareal},
\begin{equation}
\alpha^\ast s-\alpha s^\ast
=\alpha\left(\frac{i g}{\gamma_2}\alpha w\right)-\alpha\left(-\frac{i g}{\gamma_2}\alpha w\right)
=\frac{2 i g}{\gamma_2}\,\alpha^2 w.
\end{equation}
Inserting this into $\dot w=0$ in Eq.~\eqref{app:eq_w} gives
\begin{equation}
0=-\gamma_1(w-S_z^{(0)})-\frac{4g^2}{\gamma_2}\alpha^2 w,
\label{app:w_equation}
\end{equation}
or equivalently
\begin{equation}
w(\alpha)=\frac{S_z^{(0)}}{1+\displaystyle \frac{4g^2}{\gamma_1\gamma_2}\alpha^2 }.
\label{app:w_of_alpha}
\end{equation}
This explicitly exhibits how stimulated emission depletes the inversion as the coherent mechanical field builds up.

\paragraph{Step 3: gain clamping and the nontrivial steady state.}
From $\dot\alpha=0$ in Eq.~\eqref{app:eq_alpha}, together with Eq.~\eqref{app:s_solution},
\begin{equation}
0=-\frac{\gamma_m}{2}\alpha - i g\left(\frac{i g}{\gamma_2}\alpha w\right)
=-\frac{\gamma_m}{2}\alpha+\frac{g^2}{\gamma_2}\alpha w.
\label{app:alpha_equation}
\end{equation}
Besides the trivial solution $\alpha=0$ (below threshold), a nonzero steady state requires
\begin{equation}
w=\frac{\gamma_m\gamma_2}{2g^2}\equiv S_z^{(\mathrm{th})},
\label{app:w_clamp}
\end{equation}
which is the standard gain-clamping condition: above threshold, the inversion adjusts to the threshold value so that gain exactly balances loss. (The identification with $S_z^{(\mathrm{th})}$ uses the threshold derived in Eq.~\eqref{app:Szth}.)

Combining Eq.~\eqref{app:w_clamp} with Eq.~\eqref{app:w_of_alpha} yields
\begin{equation}
S_z^{(\mathrm{th})}
=\frac{S_z^{(0)}}{1+\displaystyle \frac{4g^2}{\gamma_1\gamma_2}\alpha^2 }
\quad\Longrightarrow\quad
\alpha^2
=\frac{\gamma_1\gamma_2}{4g^2}\left(\frac{S_z^{(0)}}{S_z^{(\mathrm{th})}}-1\right),
\label{app:alpha2}
\end{equation}
with $S_z^{(0)}>S_z^{(\mathrm{th})}$. Identifying the coherent phonon number with $n_{\mathrm{las}}\equiv|\alpha|^2=\alpha^2$ (in this mean-field treatment), we obtain
\begin{equation}
n_{\mathrm{las}}
= n_{\mathrm{sat}}\left(\frac{S_z^{(0)}}{S_z^{(\mathrm{th})}}-1\right),
\qquad
n_{\mathrm{sat}}\equiv \frac{\gamma_1\gamma_2}{4g^2}.
\label{app:nlas}
\end{equation}
Equation~\eqref{app:nlas} is the Maxwell--Bloch steady-state result: the coherent phonon population grows linearly with the excess inversion and is suppressed by stronger coupling (larger $g$) through saturation.

\paragraph{Consistency check with the linear threshold.}
Linearizing Eqs.~\eqref{app:eq_alpha}--\eqref{app:eq_w} around $\alpha=0$ and $w\simeq S_z^{(0)}$ gives
\begin{equation}
\dot\alpha \simeq \left(-\frac{\gamma_m}{2}+\frac{g^2}{\gamma_2}S_z^{(0)}\right)\alpha,
\end{equation}
so the amplitude growth rate changes sign at $S_z^{(0)}=\gamma_m\gamma_2/(2g^2)$, in agreement with Eq.~\eqref{app:Szth} derived from the reduced master equation.

\subsection{Order-of-magnitude estimates with parameters from Ref.~\cite{hatifi2025}}
\label{app:estimates}

We now translate the analytical gain, threshold, and saturation laws derived previously into concrete numbers using the representative parameter set of \cite{hatifi2025}. Throughout, $g$ denotes the \emph{effective} transverse Jaynes--Cummings coupling entering $H_{\rm JC}$ Eq.~\eqref{app:HJC}, i.e.\ after dressing and the near-resonant RWA.

\paragraph{Input parameters and conventions.}
We use angular frequencies and rates in ${\rm s^{-1}}$. The mechanical damping is $\gamma_m=\omega_m/Q$. The dressed-spin rates are expressed through $(\gamma_1,\gamma_2)$ as in Eq.~\eqref{app:gamma12}. To connect directly to the coupling window discussed in Ref.~\cite{hatifi2025}, we parametrize $g=\eta\,\omega_m$ and consider $\eta=0.05,\,0.10$ as representative, with $\eta=0.25$ as a strong-coupling benchmark. For convenience, Table~\ref{tab:est_inputs} collects the numerical inputs and the directly inferred mechanical damping rate used throughout this subsection.

\paragraph{Threshold inversion and maximum small-signal gain.}
\begin{table}[t]
\caption{Representative parameters used for the estimates in this Appendix subsection (from Table~I of Ref.~\cite{hatifi2025}). Frequencies are given as angular frequencies; all dissipative rates are in ${\rm s^{-1}}$.}
\label{tab:est_inputs1}
\begin{ruledtabular}
\begin{tabular}{lcc}
Quantity & Value & Unit\\
\hline
Mechanical frequency & $\omega_m=2\pi\times 50$ & ${\rm s^{-1}}$\\
Quality factor & $Q=10^{4}$ & --\\
Mechanical damping & $\gamma_m=\omega_m/Q\simeq 3.14\times 10^{-2}$ & ${\rm s^{-1}}$\\
Spin relaxation & $\gamma_1=5\times 10^{2}$ & ${\rm s^{-1}}$\\
Spin decoherence & $\gamma_2=10^{3}$ & ${\rm s^{-1}}$\\
\end{tabular}
\end{ruledtabular}
\end{table}
\begin{table}[t]
\caption{Derived gain and saturation benchmarks versus the dimensionless coupling $g/\omega_m=\eta$, using the inputs of Table~\ref{tab:est_inputs} and the on-resonance formulas.}
\label{tab:eta_summary1}
\begin{ruledtabular}
\begin{tabular}{lccc}
& $\eta=0.05$ & $\eta=0.10$ & $\eta=0.25$ \\
\hline
$g$ [$\mathrm{s^{-1}}$] & $1.57\times 10^{1}$ & $3.14\times 10^{1}$ & $7.85\times 10^{1}$ \\
$S_z^{(\mathrm{th})}$ & $6.37\times 10^{-2}$ & $1.59\times 10^{-2}$ & $2.55\times 10^{-3}$ \\
$|\Gamma_{\mathrm{NV}}|_{\max}=2g^2/\gamma_2$ [$\mathrm{s^{-1}}$] &
$4.93\times 10^{-1}$ & $1.97$ & $1.23\times 10^{1}$ \\
$|\Gamma_{\mathrm{NV}}|_{\max}/\gamma_m$ & $1.57\times 10^{1}$ & $6.27\times 10^{1}$ & $3.92\times 10^{2}$ \\
$n_{\rm sat}=\gamma_1\gamma_2/(4g^2)$ & $5.07\times 10^{2}$ & $1.27\times 10^{2}$ & $2.03\times 10^{1}$ \\
\end{tabular}
\end{ruledtabular}
\end{table}

On resonance ($\delta=0$), the spin-induced contribution to the phonon-number damping is $\Gamma_{\mathrm{NV}}=-\frac{2g^2}{\gamma_2}S_z^{(0)}$, so the threshold condition $\gamma_{\rm eff}^{(n)}=\gamma_m+\Gamma_{\mathrm{NV}}=0$ yields
\begin{equation}
S_z^{(\mathrm{th})}=\frac{\gamma_m\gamma_2}{2g^2}
=\frac{\gamma_2}{2\eta^2\,\omega_m Q}.
\label{app:Szth_eta}
\end{equation}
A useful gain benchmark is the maximum magnitude of negative number damping achievable at full dressed inversion ($S_z^{(0)}=1$),
\begin{equation}
|\Gamma_{\mathrm{NV}}|_{\max}=\frac{2g^2}{\gamma_2}.
\label{app:Gammaopt_max_def}
\end{equation}
Table~\ref{tab:eta_summary1} summarizes $S_z^{(\mathrm{th})}$, $|\Gamma_{\mathrm{NV}}|_{\max}$, and the gain margin $|\Gamma_{\mathrm{NV}}|_{\max}/\gamma_m$ for the three values of $\eta$. These numbers show that, within the parameter window of Ref.~\cite{hatifi2025}, the threshold inversion is at the percent level for $\eta\gtrsim 0.1$, while even for $\eta=0.05$ the gain margin at full inversion exceeds intrinsic mechanical loss by more than an order of magnitude. 
\paragraph{Saturation scale and above-threshold coherent phonon number.}
In the mean-field saturation theory of the previous section, the inversion is depleted until gain clamps at threshold, and the steady-state coherent phonon number obeys
\begin{equation}
n_{\mathrm{las}}
= n_{\rm sat}\!\left(\frac{S_z^{(0)}}{S_z^{(\mathrm{th})}}-1\right),
\qquad
n_{\rm sat}=\frac{\gamma_1\gamma_2}{4g^2}.
\label{app:nlas_recall}
\end{equation}
Eliminating $S_z^{(\mathrm{th})}$ using Eq.~\eqref{app:Szth_eta} gives the equivalent form
\begin{equation}
n_{\mathrm{las}}
=\frac{\gamma_1}{2\gamma_m}\,S_z^{(0)}-n_{\rm sat},
\qquad (\delta=0),
\label{app:nlas_simplified}
\end{equation}
which makes explicit that the dominant scale is $\gamma_1/\gamma_m$ (spin repumping versus mechanical energy decay), while $g$ primarily controls the offset $n_{\rm sat}$ reported in Table~\ref{tab:eta_summary1}. For the inputs of Table~\ref{tab:est_inputs1}, $\gamma_1/(2\gamma_m)\simeq 7.96\times 10^{3}$, so moderate inversion already yields a sizeable coherent population above threshold.

\paragraph{Thermal occupation and the ultra-low-frequency constraint.}
The reduced master equation includes an intrinsic thermal bath with mean occupation $\bar n_{\rm th}$. In the high-temperature limit $\hbar\omega_m\ll k_BT$,
\begin{equation}
\bar n_{\rm th}\simeq \frac{k_BT}{\hbar\omega_m}.
\label{app:nth_highT}
\end{equation}
For $\omega_m=2\pi\times 50~{\rm s^{-1}}$,
\begin{equation}
\bar n_{\rm th}\simeq 1.3\times 10^{11}\left(\frac{T}{300~{\rm K}}\right).
\label{app:nth_50Hz}
\end{equation}
Thus, at ambient temperature the total phonon population is dominated by thermal noise. In that regime, the present single-spin mechanism can still generate \emph{negative damping} and self-oscillation, but the coherent phonon population predicted by Eq.~\eqref{app:nlas_simplified} remains negligible compared to $\bar n_{\rm th}$ unless the \emph{effective} motional temperature is strongly reduced. A conservative visibility criterion is $n_{\mathrm{las}}\gtrsim \bar n_{\rm th}$, implying
\begin{equation}
T_{\rm eff}\lesssim \frac{\hbar\omega_m}{k_B}\,n_{\mathrm{las}}
\simeq 1.9\times 10^{-5}~{\rm K}\left(\frac{n_{\mathrm{las}}}{8\times 10^{3}}\right),
\label{app:Teff_req}
\end{equation}
for the present $50$~Hz oscillator. This motivates two practical routes to enhance observability of phonon-lasing signatures: (i) increasing $\omega_m$ (thereby suppressing $\bar n_{\rm th}$ at fixed $T$), and/or (ii) reducing the effective motional temperature via feedback/sideband cooling before (or during) maser operation. In either case, the gain and threshold expressions derived above remain directly applicable after updating $\omega_m$, $\gamma_m$, and the effective $g$ to the relevant platform.
\subsection{Semiclassical Langevin description and phase-space ring steady state}
\label{app:langevin}

The full master-equation provides a microscopic validation of the maser regime but become heavy when the mechanical Hilbert space must accommodate large steady-state occupations. For the purpose of producing physically transparent phase-space figures and extracting standard maser diagnostics, it is useful to employ a semiclassical drift--diffusion description of the mechanical mode that is consistent with the reduced master equation \eqref{eq:MEred} and with the Maxwell--Bloch saturation law \eqref{app:nlas_recall}. In this Appendix, we develop the Langevin model used to generate the phase-diffusing ring and related diagnostics.

\paragraph{From the reduced Lindblad equation to a Wigner Fokker--Planck equation.}
The reduced mechanical master equation after adiabatic elimination, has the standard form
\begin{equation}
\begin{aligned}
\dot{\rho}_m
=
-\frac{i}{\hbar}&\Big[\hbar(\omega_m+\delta\omega)\,a^\dagger a,\rho_m\Big] \nonumber \\
&+ A_-\,\mathcal{D}[a]\rho_m
+ A_+\,\mathcal{D}[a^\dagger]\rho_m .
\end{aligned}
\label{app:MEred_Apm}
\end{equation}
with total downward/upward transition rates
\begin{equation}
A_- \equiv \gamma_m(\bar n_{\rm eff}+1)+\Gamma_-(\delta),
\quad
A_+ \equiv \gamma_m\bar n_{\rm eff}+\Gamma_+(\delta),
\label{app:Apm_def_again}
\end{equation}
where $\bar n_{\rm eff}$ denotes the effective mechanical bath occupation at the operating point (e.g., after pre-cooling or feedback), and $\Gamma_\pm(\delta)$ are given by Eq.~\eqref{eq:Gammapm}. The Wigner function $W(z,p,t)$ of the mechanical mode obeys a Fokker--Planck equation with linear drift and isotropic diffusion for such a Lindblad generator. Writing dimensionless quadratures
\begin{equation}
z \equiv \frac{a+a^\dagger}{\sqrt{2}},\qquad
p \equiv \frac{a-a^\dagger}{i\sqrt{2}},
\label{app:xp_def}
\end{equation}
so that $n=a^\dagger a = (z^2+p^2-1)/2$ at the operator level and $n\simeq (z^2+p^2)/2$ in the large occupation regime, the dissipators produce a diffusion constant
\begin{equation}
D_{\rm W}=\frac{A_-+A_+}{4}
=\frac{\gamma_m(2\bar n_{\rm eff}+1)+\Gamma_-(\delta)+\Gamma_+(\delta)}{4}.
\label{app:DW_def}
\end{equation}
Using Eq.~\eqref{eq:Gammapm} one may also write $\Gamma_-(\delta)+\Gamma_+(\delta)=2g^2\gamma_2/(\gamma_2^2+\delta^2)\equiv 2G(\delta)$ with
\begin{equation}
G(\delta)\equiv \frac{g^2\gamma_2}{\gamma_2^2+\delta^2}.
\label{app:G_def}
\end{equation}
In the rotating frame where the residual Hamiltonian is $\hbar\Omega_m a^\dagger a$ (with $\Omega_m$ possibly including small frequency shifts), the corresponding Wigner Fokker--Planck equation reads
\begin{equation}
\begin{aligned}
\partial_t W={}&-\partial_z\!\big[(\mu \,z+\Omega_m p)W\big]
-\partial_p\!\big[(\mu\, p-\Omega_m z)W\big] \\
&\quad + D_{\rm W}(\partial_z^2+\partial_p^2)W ,
\end{aligned}
\label{app:FP_linear}
\end{equation}
where $\mu$ is the effective (linear) amplitude drift coefficient. In a strictly linear theory, $\mu=-(A_- -A_+)/2$ and the steady state is Gaussian. In the maser regime, however, gain saturates and the drift becomes amplitude dependent, as described next.

\paragraph{Nonlinear drift from Maxwell--Bloch gain saturation.} 
Above threshold, stimulated emission reduces the dressed inversion and clamps the net gain. A Maxwell--Bloch mean-field elimination yields an inversion depletion of the form
\begin{equation}
w(n)\equiv \langle\tilde\sigma_z\rangle \simeq \frac{S_z^{(0)}}{1+n/n_{\rm sat}},
\quad
n_{\rm sat}=\frac{\gamma_1}{4G(\delta)}=\frac{\gamma_1(\gamma_2^2+\delta^2)}{4g^2\gamma_2},
\label{app:w_of_n}
\end{equation}
where $n\simeq|\alpha|^2$ is the (classical) phonon number associated with the complex amplitude $\alpha\equiv \langle a\rangle$, and $S_z^{(0)}$ is the small-signal inversion \eqref{app:Sz0}. In the same spirit, the signal gain coefficient entering the amplitude equation is $G(\delta)$ in Eq.~\eqref{app:G_def}. The resulting nonlinear drift for the mechanical amplitude is then
\begin{equation}
\mu(n)= -\frac{\gamma_m}{2} + G(\delta)\,w(n)
= -\frac{\gamma_m}{2}+G(\delta)\,\frac{S_z^{(0)}}{1+n/n_{\rm sat}}.
\label{app:mu_of_n}
\end{equation}
The linear threshold condition follows from $\mu(0)=0$, yielding $S_z^{(\mathrm{th})}(\delta)=\gamma_m/[2G(\delta)]$, consistent with Eq.~\eqref{eq:Szth} and Eq.~\eqref{app:Szth_detuned}. In the noiseless limit, the nontrivial steady-state solution $\mu(n_{\rm las})=0$ reproduces the Maxwell--Bloch coherent occupation
\begin{equation}
n_{\rm las}
=
n_{\rm sat}\!\left(\frac{S_z^{(0)}}{S_z^{(\mathrm{th})}}-1\right),
\qquad S_z^{(0)}>S_z^{(\mathrm{th})},
\label{app:nlas_again}
\end{equation}
which sets the predicted ring radius in phase space.

\paragraph{Equivalent Langevin equations.}
Equation~\eqref{app:FP_linear} with the nonlinear drift $\mu(n)$ in Eq.~\eqref{app:mu_of_n} is equivalent to the following It\^o Langevin system for the real quadratures $(z,p)$:
\begin{align}
dz&= (\mu(n)\,z+\Omega_m\, p)\,dt + \sqrt{2D_{\rm W}}\,d\xi_z,
\label{app:Langevin_x}\\
dp &= (\mu(n)\,p-\Omega_m z)\,dt + \sqrt{2D_{\rm W}}\,d\xi_p,
\label{app:Langevin_p}
\end{align}
where $d\xi_z$ and $d\xi_p$ are independent Wiener increments with $\langle d\xi_i\rangle=0$ and $\langle d\xi_i(t)\,d\xi_j(t)\rangle=\delta_{ij}\,dt$. In our numerics we integrate Eqs.~\eqref{app:Langevin_x}--\eqref{app:Langevin_p} using a standard Euler--Maruyama scheme and construct steady-state phase-space histograms from long-time samples.

\paragraph{Phase-diffusing ring and the distinction between $n_{\rm las}$ and $\langle n\rangle$.}
Above threshold, the deterministic dynamics $\dot z=\mu(n)z+\Omega_m\, p$ and $\dot p=\mu(n)p-\Omega_m\, z$ possesses a stable limit cycle at radius $r_0=\sqrt{z^2+p^2}=\sqrt{2n_{\rm las}}$, while the global phase is neutrally stable. The additive isotropic diffusion in Eqs.~\eqref{app:Langevin_x}--\eqref{app:Langevin_p} therefore produces a steady state that is uniform in phase and peaked in radius, i.e.\ a \emph{ring} in $(z,p)$ space. Importantly, the Maxwell--Bloch prediction $n_{\rm las}$ characterizes the deterministic limit-cycle intensity (coherent component), but the Langevin steady state includes additional incoherent fluctuations generated by diffusion. Because $n=(z^2+p^2)/2$ is quadratic in fluctuating variables, diffusion biases the mean upward:
\begin{equation}
\langle n\rangle_{\rm ss}
=
\frac{\langle z^2+p^2\rangle_{\rm ss}}{2}
=
n_{\rm las}
+\frac{\mathrm{Var}(z)+\mathrm{Var}(p)}{2},
\label{app:n_mean_bias}
\end{equation}
so that generically $\langle n\rangle_{\rm ss}>n_{\rm las}$ which when $(\mathrm{Var}(z)+\mathrm{Var}(p))/2\ll n_{\rm las}$ gives a radius close to $\sqrt{2n_{\rm las}}$. This explains why time traces of $\langle n(t)\rangle$ saturate at a value slightly larger than $n_{\rm las}$ when diffusion is included.

\paragraph{Closed-form stationary radial distribution (Fokker--Planck solution).}
Because the drift and diffusion are rotationally symmetric, the steady state depends only on $r=\sqrt{z^2+p^2}$. The stationary Fokker--Planck equation admits the (unnormalized) solution
\begin{equation}
P_{\rm ss}(z,p)\propto
\exp\!\left[\frac{1}{D_{\rm W}}\int_0^{r}\mu(s)\,s\,ds\right],
\label{app:Pss_xy}
\end{equation}
leading to a radial probability density
\begin{equation}
P_r(r)
\propto
r\,\exp\!\left[\frac{1}{D_{\rm W}}\int_0^{r}\mu(s)\,s\,ds\right],
\quad
\int_0^\infty P_r(r)\,dr=1.
\label{app:Pr_solution}
\end{equation}
From $P_r(r)$ one may compute the steady-state mean occupation,
\begin{equation}
\langle n\rangle_{\rm FP}
=
\int_0^\infty \frac{r^2}{2}\,P_r(r)\,dr,
\label{app:n_FP_mean}
\end{equation}
which accurately matches the long-time plateau of $\langle n(t)\rangle$ obtained from the Langevin simulation, thereby providing a quantitative noise-corrected counterpart to the deterministic MB prediction $n_{\rm las}$.

\paragraph{Wigner function from Langevin sampling.}
To visualize the phase-space steady state, we build a histogram of the stationary phase-space density $P_{\rm ss}(z,p)$ from Langevin samples. In the large-occupation semiclassical regime the Wigner function is well approximated by a mixture of coherent-state Wigners centered at sampled phase-space points, i.e.\ by convolving the classical density with the coherent-state kernel:
\begin{equation}
W(z,p)\simeq
\iint dz'\,dp'\,P_{\rm ss}(z',p')\,\frac{1}{\pi}
\exp\!\left[-(z-z')^2-(p-p')^2\right],
\label{app:Wigner_conv}
\end{equation}
which smooths the histogram on the natural vacuum scale. This construction captures the formation of the ring and its radius in a numerically inexpensive way and suffices for the qualitative Wigner plots reported in the main text.

\paragraph{Intensity correlations.}
Finally, the equal-time intensity correlation $g^{(2)}(0)$ is computed directly from stationary samples of
$n\simeq (z^2+p^2)/2$ as
\begin{equation}
g^{(2)}(0)=\frac{\langle n(n-1)\rangle}{\langle n\rangle^2}
=\frac{\langle n^2\rangle-\langle n\rangle}{\langle n\rangle^2}.
\label{app:g2_from_samples}
\end{equation}
Across threshold the model exhibits the standard transition from thermal-like statistics, $g^{(2)}(0)\approx 2$, to near-Poissonian statistics, $g^{(2)}(0)\to 1$, reflecting suppression of relative intensity fluctuations as the maser field becomes macroscopic.

\section{Practical design window for an \emph{observable} phonon maser}
\subsection{Interplay of $\omega_m$, $Q$, and effective temperature}
\label{app:design_window}

The previous sections establish (i) the signal threshold condition $S_z^{(0)}>S_z^{(\mathrm{th})}$ and (ii) the above-threshold coherent phonon population set by saturation Eq.~\eqref{app:nlas}. In practice, however, demonstrating \emph{phonon lasing} requires more than negative damping: the generated coherent component must be distinguishable from thermal motion. Here we make this requirement explicit and derive a useful "design window'' that clarifies what is gained by increasing $\omega_m$ and/or by engineering a lower effective motional temperature.

\paragraph{Operational criterion for observability.}
A minimal and platform-agnostic criterion is that the coherent component dominates the thermal fluctuations in the same mechanical mode. In the present mean-field picture the coherent phonon number is $n_{\mathrm{las}}=|\alpha|^2$ Eq.~\eqref{app:nlas}, while the residual incoherent occupation is set by an effective bath occupation $\bar n_{\rm eff}$ entering the mechanical Liouvillian (cf.\ Eq.~\eqref{app:Lm}). A conservative condition for observing a narrowband self-oscillation signal on top of thermal noise is therefore
\begin{equation}
n_{\mathrm{las}} \gtrsim \bar n_{\rm eff}.
\label{app:obs_criterion}
\end{equation}
In the high-temperature limit $\hbar\omega_m\ll k_BT_{\rm eff}$, one may parameterize $\bar n_{\rm eff}\simeq k_B T_{\rm eff}/(\hbar\omega_m)$, where $T_{\rm eff}$ is an \emph{effective} motional temperature (after any pre-cooling or feedback cooling).

\paragraph{From saturation to a bound on $T_{\rm eff}$.}
On resonance, Eq.~\eqref{app:Szth} together with Eq.~\eqref{app:nlas} can be rewritten as
\begin{equation}
n_{\mathrm{las}}
=\frac{\gamma_1}{2\gamma_m}\,S_z^{(0)}-n_{\rm sat},
\qquad
n_{\rm sat}\equiv \frac{\gamma_1\gamma_2}{4g^2},
\label{app:nlas_rewrite}
\end{equation}
which makes it transparent that large coherent amplitudes require a large ratio $\gamma_1/\gamma_m$ (repumping toward the uncoupled dressed steady state versus mechanical energy decay), while dephasing enters through $n_{\rm sat}$. Combining Eqs.~\eqref{app:obs_criterion} and \eqref{app:nlas_rewrite} yields an upper bound on the effective temperature for which the coherent component can dominate,
\begin{equation}
T_{\rm eff} \lesssim \frac{\hbar\omega_m}{k_B}\left(\frac{\gamma_1}{2\gamma_m}S_z^{(0)}-n_{\rm sat}\right),
\qquad (\delta=0).
\label{app:Teff_bound_general}
\end{equation}
Using $\gamma_m=\omega_m/Q$, this becomes
\begin{equation}
\begin{aligned}
T_{\rm eff} \lesssim \frac{\hbar}{k_B}
& \left(
\frac{\gamma_1 Q}{2}\,S_z^{(0)}-\omega_m n_{\rm sat}
\right)\\
&= \frac{\hbar}{k_B}
\left(
\frac{\gamma_1 Q}{2}\,S_z^{(0)}-\frac{\gamma_1\gamma_2}{4}\frac{\omega_m}{g^2}
\right),
\label{app:Teff_bound_Q}
\end{aligned}
\end{equation}
which isolates the two competing effects: increasing $Q$ and $\gamma_1$ increases the admissible $T_{\rm eff}$ linearly, whereas insufficient coupling (large $\omega_m/g^2$) reduces it. A particularly relevant limit for scaling arguments is the case where the effective JC coupling tracks the mechanical frequency, $g=\eta\,\omega_m$. In that case,
\begin{equation}
T_{\rm eff} \lesssim \frac{\hbar}{k_B}
\left(
\frac{\gamma_1 Q}{2}\,S_z^{(0)}-\frac{\gamma_1\gamma_2}{4\eta^2\,\omega_m}
\right),
\label{app:Teff_bound_eta}
\end{equation}
so the dominant constraint is the $\omega_m$-independent scale $(\hbar/k_B)(\gamma_1 Q/2)S_z^{(0)}$, with only a weak correction that improves as $\omega_m$ increases. In other words, increasing $\omega_m$ does suppress $\bar n_{\rm eff}$ at fixed $T_{\rm eff}$, but in the lasing regime it also tends to reduce $n_{\mathrm{las}}$ through the increase of $\gamma_m=\omega_m/Q$; the net \emph{visibility} of the coherent component is therefore primarily controlled by $\gamma_1 Q$ and by the attainable dressed inversion, rather than by $\omega_m$ alone.

\paragraph{Numerical implication for the Table~I parameter set.}
Using Table~III values $Q=10^4$ and $\gamma_1=1/T_1=5\times 10^2~\mathrm{s^{-1}}$, Eq.~\eqref{app:Teff_bound_eta} yields, at full inversion $S_z^{(0)}\simeq 1$,
\begin{equation}
T_{\rm eff}^{\max}\approx \frac{\hbar}{k_B}\,\frac{\gamma_1 Q}{2}
\simeq 1.9\times 10^{-5}~\mathrm{K},
\label{app:Teff_bound_numbers}
\end{equation}
up to the small correction term $\propto 1/\omega_m$ in Eq.~\eqref{app:Teff_bound_eta}. Equivalently, for $\omega_m=2\pi\times 50~\mathrm{rad/s}$ this corresponds to an effective occupation $\bar n_{\rm eff}\sim k_B T_{\rm eff}^{\max}/(\hbar\omega_m)\sim 8\times 10^3$, consistent with the largest coherent populations predicted by Eq.~\eqref{app:nlas_rewrite} for the same $\gamma_1/\gamma_m$.
\rem{\paragraph{Comparison with current levitated-platform capabilities.}
The bound in Eq.~\eqref{app:Teff_bound_numbers} is extremely demanding in the ultra-low-frequency regime considered here. For comparison, feedback cooling of levitated diamond particles has been demonstrated to below \(1\)~K in a magneto-gravitational trap \cite{hsu2016}, while recent high-vacuum surface-ion-trap experiments with levitated nanodiamonds have demonstrated stable operation and ODMR at pressures near \(10^{-5}\)~Torr \cite{jin2024}, but not motional cooling to the microkelvin scale required by Eq.~\eqref{app:Teff_bound_numbers}. The present benchmark should therefore be interpreted as a conservative design target for unambiguous visibility of the coherent component, rather than as a claim that this regime is already routinely accessible in present low-frequency RF levitated platforms. This also motivates the staged protocol discussed below, in which cooling and masing are applied sequentially rather than simultaneously.}

\paragraph{How pre-cooling/feedback cooling enters, and why there is a tradeoff?}
A common approach to lowering $\bar n_{\rm eff}$ is to engineer an additional cold-damping channel. At the level of the reduced description, this corresponds to supplementing Eq.~\eqref{app:Lm} by an extra Liouvillian
\begin{equation}
\mathcal{L}_{\rm cool}\rho_m
=
\gamma_{\rm cool}(\bar n_{\rm cool}+1)\mathcal{D}[a]\rho_m
+\gamma_{\rm cool}\bar n_{\rm cool}\mathcal{D}[a^\dagger]\rho_m,
\label{app:Lcool}
\end{equation}
with $\bar n_{\rm cool}\ll \bar n_{\rm th}$ for an effective cold bath. The net mechanical damping and effective occupation then become
\begin{equation}
\begin{aligned}
\gamma_m &\to \gamma_{\rm tot}\equiv \gamma_m+\gamma_{\rm cool},\\
\bar n_{\rm eff}
&=\frac{\gamma_m\bar n_{\rm th}+\gamma_{\rm cool}\bar n_{\rm cool}}{\gamma_m+\gamma_{\rm cool}}\\
&\simeq \frac{\gamma_m}{\gamma_m+\gamma_{\rm cool}}\,\bar n_{\rm th}
\quad(\bar n_{\rm cool}\simeq 0).
\end{aligned}
\label{app:neff_cooling}
\end{equation}
Crucially, however, the lasing threshold inversion depends on the \emph{total} damping, so that Eq.~\eqref{app:Szth} becomes
\begin{equation}
S_z^{(\mathrm{th})}\ \to\ S_{z,{\rm tot}}^{(\mathrm{th})}
=\frac{\gamma_{\rm tot}\gamma_2}{2g^2}
=\left(1+\frac{\gamma_{\rm cool}}{\gamma_m}\right) S_z^{(\mathrm{th})}.
\label{app:Szth_with_cooling}
\end{equation}
Equations~\eqref{app:neff_cooling} and \eqref{app:Szth_with_cooling} show the fundamental tradeoff: continuous cold damping reduces $\bar n_{\rm eff}$ but simultaneously \emph{raises} the required inversion. Indeed, in the ideal limit $\bar n_{\rm cool}\simeq 0$ their product is independent of $\gamma_{\rm cool}$,
\begin{equation}
S_{z,{\rm tot}}^{(\mathrm{th})}\,\bar n_{\rm eff}
\simeq
\frac{\gamma_m\gamma_2}{2g^2}\,\bar n_{\rm th},
\label{app:product_invariant}
\end{equation}
so cold damping does not eliminate the need for either strong coupling (large $g^2/\gamma_2$), high mechanical quality factor (small $\gamma_m$), or genuinely low motional temperature; it only allows one to trade reduced $\bar n_{\rm eff}$ against increased threshold.

A practical and experimentally natural implication is that \emph{pre-cooling and lasing need not be simultaneous}: one may first reduce the oscillator to a target $\bar n_{\rm eff}$ (using the protocol of Ref.~\cite{hatifi2025} or external feedback/sideband cooling), then switch to the lasing configuration with minimal added damping (returning to $\gamma_{\rm tot}\simeq \gamma_m$) while pumping the dressed inversion. In that staged protocol the threshold remains set by Eq.~\eqref{app:Szth} while the initial condition is improved, maximizing the coherent buildup predicted by Eq.~\eqref{app:nlas_rewrite} without paying the penalty in Eq.~\eqref{app:Szth_with_cooling}.

For fixed achievable dressed inversion $S_z^{(0)}$ and fixed spin coherence parameters $(\gamma_1,\gamma_2)$, the most effective levers for making phonon lasing \emph{unambiguously observable} are (a) increasing $Q$ (reducing $\gamma_m$), (b) increasing the effective JC coupling $g$ while keeping $\gamma_2$ low, and (c) reducing the effective motional temperature $T_{\rm eff}$ until the bound \eqref{app:Teff_bound_general} is satisfied. Increasing $\omega_m$ is beneficial insofar as it can be done without sacrificing $Q$ and $g$; otherwise its impact on visibility is limited by the concomitant increase of $\gamma_m=\omega_m/Q$ and by the platform-dependent scaling of $g$ with $\omega_m$. 

\rem{\subsection{Pressure scale associated with the benchmark mechanical damping}
\label{app:pressure_scale_benchmark}

The benchmark mechanical parameters used in the main text are
\[
\omega_m = 2\pi\times 50~\mathrm{s}^{-1},
\quad
Q = 10^4,
\quad
\gamma_m = \frac{\omega_m}{Q} \simeq 3.14\times 10^{-2}~\mathrm{s}^{-1}.
\]
It is useful to translate this damping scale into an order-of-magnitude background pressure in order to assess the vacuum regime corresponding to the benchmark quality factor. If the center-of-mass damping is dominated by residual-gas collisions and the particle lies in the free-molecular regime, the translational damping rate is approximately linear in pressure. At the order-of-magnitude level, one may write \cite{bullier2020}
\begin{equation}
\gamma_{\mathrm{gas}}
\sim
\alpha\,
\frac{P}{\rho a \,\bar c},
\qquad
\bar c=\sqrt{\frac{8k_B T}{\pi m_g}},
\label{app:gas_damping_scaling}
\end{equation}
where \(a\) is the particle radius, \(\rho\) its mass density, \(m_g\) the gas-molecule mass, \(T\) the gas temperature, and \(\alpha\) is an order-unity factor that depends on the gas--surface accommodation model. This expression is sufficient here for estimating the pressure scale associated with the benchmark damping.

Taking nitrogen at room temperature, \(\rho \simeq 3.5\times 10^3~\mathrm{kg\,m^{-3}}\) for diamond, and representative nanodiamond radii in the \(100\)--\(300\) nm range, the benchmark value
\[
\gamma_m \simeq 3.14\times 10^{-2}~\mathrm{s}^{-1}
\]
corresponds to pressures of order
\begin{equation}
P_{\mathrm{req}} \sim 10^{-5}\text{--}10^{-4}\ \mathrm{mbar},
\label{app:pressure_window_estimate}
\end{equation}
with the lower end of the interval associated with smaller particles. The quality factor \(Q=10^4\), therefore, corresponds to a high-vacuum regime, but not to an implausible pressure scale for levitated-particle platforms.

This estimate also helps separate two distinct issues. The first is the vacuum level required to reach the benchmark mechanical damping. The second is the compatibility of that regime with sustained optical pumping of an NV-hosting nanodiamond. The present effective theory addresses the first point only indirectly, through the parameter \(\gamma_m\), and does not attempt to model the full thermal and photophysical constraints associated with continuous optical illumination at low pressure. For this reason, the benchmark parameter set should be interpreted as a design-level operating point for the gain and threshold analysis. The observability of the maser signal is then not determined by \(\gamma_m\) alone, but by the combined requirements of threshold, residual thermal occupation, and practical cooling strategy. }

\end{document}